\documentclass[fleqn,usenatbib]{mnras}

\usepackage{newtxtext,newtxmath}

\usepackage[T1]{fontenc}

\DeclareRobustCommand{\VAN}[3]{#2}
\let\VANthebibliography\thebibliography
\def\thebibliography{\DeclareRobustCommand{\VAN}[3]{##3}\VANthebibliography}

\usepackage{graphicx}	
\usepackage{amsmath}	

\title[High-velocity gas in the Cygnus Loop]{A refined search for high-velocity gas in the Cygnus Loop supernova remnant}

\author[A. M. Ritchey et al.]{
Adam M. Ritchey,$^{1}$\thanks{E-mail: ritchey.astro@gmail.com}
S. R. Federman,$^{2}$
and David L. Lambert$^{3}$
\\
$^{1}$Eureka Scientific, 2452 Delmer, Suite 100, Oakland, CA 96402, USA\\
$^{2}$Department of Physics and Astronomy, University of Toledo, Toledo, OH 43606, USA\\
$^{3}$W. J. McDonald Observatory and Department of Astronomy, University of Texas at Austin, Austin, TX 78712, USA
}

\date{Accepted XXX. Received YYY; in original form ZZZ}

\pubyear{2023}

\begin{document}
\label{firstpage}
\pagerange{\pageref{firstpage}--\pageref{lastpage}}
\maketitle

\begin{abstract}
We present the results of a sensitive search for high-velocity gas in interstellar absorption lines associated with the Cygnus Loop supernova remnant (SNR). We examine high-resolution, high signal-to-noise ratio optical spectra of six stars in the Cygnus Loop region with distances greater than $\sim$700 pc. All stars show low-velocity Na~{\sc i} and Ca~{\sc ii} absorption. However, only one star, HD~198301, exhibits high-velocity Ca~{\sc ii} absorption components, at velocities of +62, +82, and +96 km~s$^{-1}$. The distance to this star of $\sim$870 pc helps to constrain the distance to the receding edge of the Cygnus Loop's expanding shock front. One of our targets, HD~335334, was previously thought to exhibit high positive and high negative velocity interstellar Na~{\sc i} and Ca~{\sc ii} absorption. This was one factor leading Fesen et al. to derive a distance to the Cygnus Loop of $725\pm15$ pc. However, we find that HD~335334 is in fact a double-line spectroscopic binary and shows no evidence of high-velocity interstellar absorption. As such, the distance to HD~335334 cannot be used to constrain the distance to the Cygnus Loop. Our detection of Ca~{\sc ii} absorption approaching 100 km~s$^{-1}$ toward HD~198301 is the first conclusive detection of high-velocity absorption from a low ionization species associated with the Cygnus Loop SNR. A large jump in the Na~{\sc i} column density toward BD+31~4218, a star located beyond the northwestern boundary of the Cygnus Loop, helps to constrain the distance to a large molecular cloud complex with which the Cygnus Loop is evidently interacting.
\end{abstract}




\section{Introduction}
The Cygnus Loop supernova remnant (SNR), also known as the Veil Nebula, is one of the best-studied evolved Galactic SNRs owing to its relative brightness, its large angular size, and the absence of a significant amount of foreground extinction along the line of sight. With an extensive network of well-resolved filamentary structures, the Cygnus Loop SNR is an excellent laboratory for investigations into various shock-related phenomena, including cloud-shock interactions \citep[e.g.,][]{d01,p02}, X-ray, UV, and optical emission from pre- and post-shock gas \citep[e.g.,][]{s00,b02,s09,m14,ray23}, and dust grain destruction via shock sputtering \citep[e.g.,][]{s10,r13}.

However, despite its relative proximity, an accurate distance to the Cygnus Loop SNR has been difficult to determine. \citet{m58} derived a distance of 770 pc from a kinematic investigation of the remnant's bright optical filaments. More recent distance estimates, mainly from proper motion studies, show considerable variation, ranging from 440 pc to 1400 pc \citep[see the summary of results provided by][]{f18a}. As with any SNR, the distance to the Cygnus Loop is a key parameter for understanding many fundamental properties of the remnant, including the shock speed, the gas pressure, and the SN explosion energy.

\citet{b09} detected high-velocity interstellar O~{\sc vi} absorption toward the subdwarf OB star KPD~2055+3111, which is positioned among the bright optical filaments in the eastern portion of the Cygnus Loop (known as the Eastern Veil Nebula). From an analysis of the star's optical spectrum, \citet{b09} obtained stellar parameters that allowed them to calculate a distance of $576\pm61$ pc to KPD~2055+3111. Since this star presumably lies behind the Cygnus Loop SNR, its distance sets a hard upper limit on the distance to the Cygnus Loop. However, Gaia EDR3 parallax measurements indicate that the distance to KPD~2055+3111 is $819^{+21}_{-18}$ pc \citep{bj21}, significantly larger than the estimate by \citet{b09}.

\citet{f18a} estimated the distance to the Cygnus Loop based on the distances to two stars they suggested are interacting with the remnant. One of the stars, a red giant with the designation TYC~2688-1037-1, is surrounded by a faint emission nebula, which \citet{f18a} proposed may be due to the interaction between the SNR blast wave and mass-loss material from the star. However, the Gaia EDR3 distance to TYC~2688-1037-1 is $2160\pm130$ pc \citep{bj21}, indicating that this star is located far behind the Cygnus Loop SNR. \citet{f18a} also observed a bow-shaped nebula near the star BD+31~4224, and suggested that a bow shock was created by the interaction between the star's stellar wind and the remnant's expanding shock front. The Gaia EDR3 distance to BD+31~4224 is $726^{+13}_{-11}$ pc \citep{bj21}. Thus, a connection with the Cygnus Loop SNR is plausible.

In a subsequent study, \citet{f18b} searched for high-velocity interstellar Na~{\sc i} and Ca~{\sc ii} absorption toward several stars in the Cygnus Loop region. \citet{f18b} claimed to discover high velocity gas toward three stars: HD~335334, TYC~2688-365-1, and TYC~2692-3378-1. Based on their discovery of high velocity gas, and adopting Gaia DR2 distances to the stars, \citet{f18b} constrained the distance to the Cygnus Loop SNR to be $735\pm25$ pc. More recently, \citet{f21} revised their distance estimate to $725\pm15$ pc, based on Gaia EDR3 measurements of the stars previously found to exhibit high-velocity interstellar absorption.

There is a problem with the conclusions of \citet{f18b}, however. These authors used low resolution spectra (with $\Delta v\approx30$$-$$45$ km~s$^{-1}$) to search for high-velocity interstellar Na~{\sc i} and Ca~{\sc ii} absorption lines. As we will show in this work, the star with the clearest signature of high-velocity absorption, HD~335334, is actually a double-line spectroscopic binary. The ``high-velocity'' components observed by \citet{f18b} toward this star are actually the stellar Na~{\sc i} D and Ca~{\sc ii} K lines from the primary and the secondary. After accounting for stellar absorption, there are no high-velocity interstellar Na~{\sc i} or Ca~{\sc ii} components toward HD~335334. The distance to this star, therefore, cannot be used to constrain the distance to the Cygnus Loop. The other two stars that \citet{f18b} claimed show high-velocity gas exhibit narrow low-velocity interstellar Na~{\sc i} components superimposed onto what appear to be broad stellar Na~{\sc i} absorption lines. (The Ca~{\sc ii}~K line in the spectrum of TYC~2692-3378-1 also appears to be very broad.) Thus, the ``high-velocity'' components in these cases are likely just the wings of the broad stellar absorption features.

A previous attempt by \citet{w02} to search for high-velocity interstellar Na~{\sc i} and Ca~{\sc ii} absorption toward stars in the vicinity of the Cygnus Loop failed to detect any absorption features with velocities greater than $\sim$30 km~s$^{-1}$ (relative to the local standard of rest; LSR). In hindsight, this failure may, at least in part, be due to the fact that most of the stars observed by \citet{w02} have Gaia EDR3 distances of less than $\sim$630 pc, and so may lie in front of the SNR. Moreover, the most distant star observed by \citet{w02}, HD~197702, lies significantly outside the boundary of the remnant, as deduced from the bright H$\alpha$ emission contours \citep[see, e.g., Figure~1 in][]{w02}, and so would not be expected to show high-velocity interstellar gas.

In this investigation, we present a new sensitive search for high-velocity interstellar absorption associated with the Cygnus Loop SNR. We examine high-resolution ($\Delta v\approx4.5$ km~s$^{-1}$), high signal-to-noise (S/N) ratio optical spectra of six stars in the Cygnus Loop region with Gaia EDR3 distances of $\sim$700 pc or greater. We detect high-velocity Ca~{\sc ii} absorption components (with LSR velocities approaching 100 km~s$^{-1}$) toward only one star: HD~198301. This is the first conclusive detection of high-velocity Ca~{\sc ii} absorption associated with the Cygnus Loop. Our observations and data reduction procedures are described in Section 2. In Section 3.1, we explain how we carefully separated stellar absorption lines from interstellar absorption features for our targets. In Section 3.2, we describe the profile fitting routine used to obtain interstellar column densities and component structures. We compare our survey with that of \citet{w02} in Section 3.3. The implications of our results for distance estimates to the Cygnus Loop SNR are discussed in Section 4. Our main conclusions are presented in Section 5. In Appendix A, we give a brief explanation of the procedures used to derive spectral types and luminosity classes for our program stars. (Prior to our study, most of our targets had only limited information available concerning their spectral classification.)

\section{Observations and Data Reduction}
\begin{figure}
\includegraphics[width=\columnwidth]{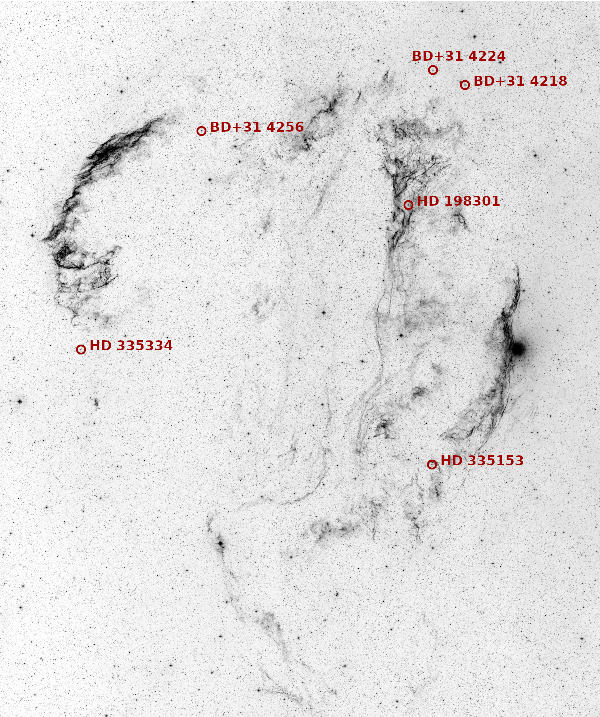}
\caption{Optical image of the Cygnus Loop supernova remnant from the Digitized Sky Survey (DSS2 red). The six stars targeted for our observing program are indicated.}
\label{fig:cygnus_dss2}
\end{figure}

Six stars were observed using the Tull spectrograph \citep[TS23;][]{t95} on the 2.7 m Harlan J.~Smith Telescope at McDonald Observatory over the course of six nights in 2022 September. The targets were selected based on several criteria. Each potential target was required to have a Gaia EDR3 distance of 700 pc or greater and to be positioned within or very near the optical boundary of the Cygnus Loop SNR. We then selected stars with $B$ and $V$ magnitudes less than 10 so that high S/N ratio optical spectra could be acquired within a reasonable amount of time. The on-sky positions of the six targets in relation to the optical nebulosities associated with the Cygnus Loop are shown in Figure~\ref{fig:cygnus_dss2}. Note that our target list includes BD+31~4224, which \citet{f18a} suggested is interacting with the Cygnus Loop, and HD~335334, which \citet{f18b} claimed shows high-velocity interstellar absorption. We had planned on obtaining data for a seventh target, HD~335249. However, due to adverse weather conditions at the beginning of the run, we decided to eliminate this relatively faint star from the target list.

Basic information regarding the target stars is provided in Table~\ref{tab:targets}. The coordinates and $B$ and $V$ magnitudes are from the SIMBAD database \citep{w00}. The spectral types listed were derived in this work (see Appendix A). The values of $E$($B$$-$$V$) were determined based on the derived spectral types, using intrinsic colors from \citet{we94}. The distances provided in Table~\ref{tab:targets} are those derived from Gaia EDR3 parallax measurements \citep{bj21}.

\begin{table*}
\centering
\caption{Stellar and observational data for the program stars. The spectral types listed are those derived in this work (see Appendix A). Stellar coordinates and $B$ and $V$ magnitudes are from the SIMBAD database \citep{w00}. Values of $E$($B$$-$$V$) were determined using intrinsic colors from \citet{we94}. The distances provided are from Gaia EDR3 parallax measurements \citep{bj21}. The last column gives the total exposure time on each target.}
\label{tab:targets}
\begin{tabular}{lcccccccc}
\hline
Star & Sp.~Type & R.~A. & Dec. & $B$ & $V$ & $E$($B$$-$$V$) & $d$ & Exp.~Time \\
 & & (J2000) & (J2000) & (mag) & (mag) & (mag) & (pc) & (s) \\
\hline
BD+31~4218 & B2 IVe & 20 47 01.93 & +32 09 26.6 & 8.93 & 8.80 & 0.33 & $1121^{+22}_{-19}$ & 7200 \\
HD~335153 & B9 III-IV & 20 47 49.67 & +30 05 31.0 & 9.56 & 9.56 & 0.07 & $700^{+11}_{-11}$ & 10800 \\
BD+31~4224 & B7 V & 20 47 51.82 & +32 14 11.4 & 9.53 & 9.58 & 0.08 & $726^{+13}_{-11}$ & 10800 \\
HD~198301 & B8 IV & 20 48 28.26 & +31 30 11.2 & 8.62 & 8.66 & 0.06 & $872^{+21}_{-20}$ & 16200 \\
BD+31~4256 & B9 III-IV & 20 53 47.40 & +31 53 28.4 & 9.23 & 9.24 & 0.06 & $948^{+17}_{-20}$ & 9000 \\
HD~335334 & B9 V & 20 56 44.63 & +30 41 14.4 & 9.54 & 9.51 & 0.10 & $732^{+12}_{-11}$ & 14400 \\
\hline
\end{tabular}
\end{table*}

The TS23 configuration of the Tull spectrograph, when combined with a $2048\times2048$ CCD, provides nearly complete wavelength coverage of the optical spectrum, from $\sim$3700$-$10,100~\AA{}, at a nominal resolving power of $R=60,000$. Exposure times were calculated to yield S/N ratios of 100$-$200 at Ca~{\sc ii}~K. The last column of Table~\ref{tab:targets} gives the total exposure time achieved for each target. Individual exposures were limited to 30 minutes to minimize the effects of cosmic ray hits during the integrations. Standard calibration exposures (biases and flats) were obtained at the beginning of each night, while Th-Ar comparison lamp exposures were obtained throughout the night at intervals of 2$-$3 hours. From the widths of Th~{\sc i} emission lines in the comparison spectra, we find that the actual resolving power during our observing run was $R\approx66,000$, corresponding to a velocity resolution of $\sim$4.5 km~s$^{-1}$. To aid in the removal of telluric absorption lines, particularly near the Na~{\sc i}~D doublet and the K~{\sc i}~$\lambda7698$ line, a bright, unreddened star (e.g., $\alpha$~Peg) was observed each night along with the science targets.

The raw science exposures were reduced following standard procedures within the IRAF environment. The average bias frame was subtracted from the flats and science exposures and from the comparison lamp frames. Cosmic rays were then removed from the science and Th-Ar exposures. Cosmic rays were effectively removed from the flats by taking the median of all the flats obtained on a given night. Scattered light was subtracted from the median flat and from the science exposures in both the dispersion and cross-dispersion directions. Unfortunately, the flat lamp in use with the Tull spectrograph exhibits emission features at the locations of the Na~{\sc i}~D lines. Thus, we extracted one-dimensional spectra from the scattered-light corrected flats in order to remove these emission features. One-dimensional spectra were also extracted from the science and Th-Ar exposures, and the corrected flat spectra were divided into the science and comparison lamp spectra. We also performed a traditional two-dimensional flat-fielding to check that there was essentially no difference between the two procedures (outside the regions affected by the Na~{\sc i} emission features).

Wavelength solutions were obtained by identifying emission lines in the Th-Ar comparison spectra. After applying the wavelength solutions to the science spectra, the next step is to correct for telluric absorption near the Na~{\sc i} and K~{\sc i} lines. A template for telluric absorption was created from our observations of the unreddened standard star. This template was then divided into the science spectra, after correcting for differences in airmass and for small shifts in the dispersion direction. This procedure is very effective at removing relatively weak telluric absorption features from the regions surrounding the Na~{\sc i} D$_1$ and D$_2$ lines and the K~{\sc i}~$\lambda7698$ line. The stronger member of the K~{\sc i} doublet at 7664.9~\AA{} coincides with a strong atmospheric O$_2$ absorption line and therefore could not be recovered. Lastly, the science spectra were shifted to the LSR frame of reference and the multiple exposures of a given target were co-added to produce final high S/N ratio spectra. Typical S/N ratios are $\sim$150 at Ca~{\sc ii}~K and $\sim$215 at Na~{\sc i}~D.

\section{Analysis}
\subsection{Separating stellar from interstellar absorption}
For many of our targets, the process of separating stellar absorption from interstellar absorption was straightforward. BD+31~4218 is an early B-type (emission-line) star with a projected rotational velocity of $\sim$300 km~s$^{-1}$. Three other targets, HD~335153, BD+31~4224, and BD+31~4256, are late B dwarfs or subgiants with rotationally broadened absorption lines. These stars have $v \sin i\gtrsim200$ km~s$^{-1}$ (see Appendix A). In each of these cases, narrow interstellar absorption lines (e.g., Ca~{\sc ii}~H and K) are superimposed onto very broad stellar absorption features. Thus, the interstellar absorption profiles were normalized simply by fitting low-order Legendre polynomials to the smoothly varying stellar spectra.

Two of our targets, HD~198301 and HD~335334, exhibit narrow stellar absorption lines (with $v \sin i\approx20$ km~s$^{-1}$). Thus, the process of separating stellar absorption from interstellar absorption was much more difficult in these cases. This difficulty is what prompted us to attempt to derive accurate spectral classifications for our program stars. At the beginning of our investigation, the spectral types listed in SIMBAD for four of our targets (HD~335153, HD~198301, BD+31~4256, and HD~335334) were ``A0'' (with no luminosity class given). BD+31~4218 had the classification ``B2''. \citet{f18a} classified BD+31~4224 as a B7 IV-V star. We sought to improve upon these classifications so that we could correctly identify any stellar absorption that may be impacting the interstellar lines, particularly toward HD~198301 and HD~335334. Most of the details regarding our derivations of spectral types and luminosity classes for our program stars are presented in Appendix A. Here, we focus specifically on the results for HD~198301 and HD~335334.

\begin{figure}
\includegraphics[width=\columnwidth]{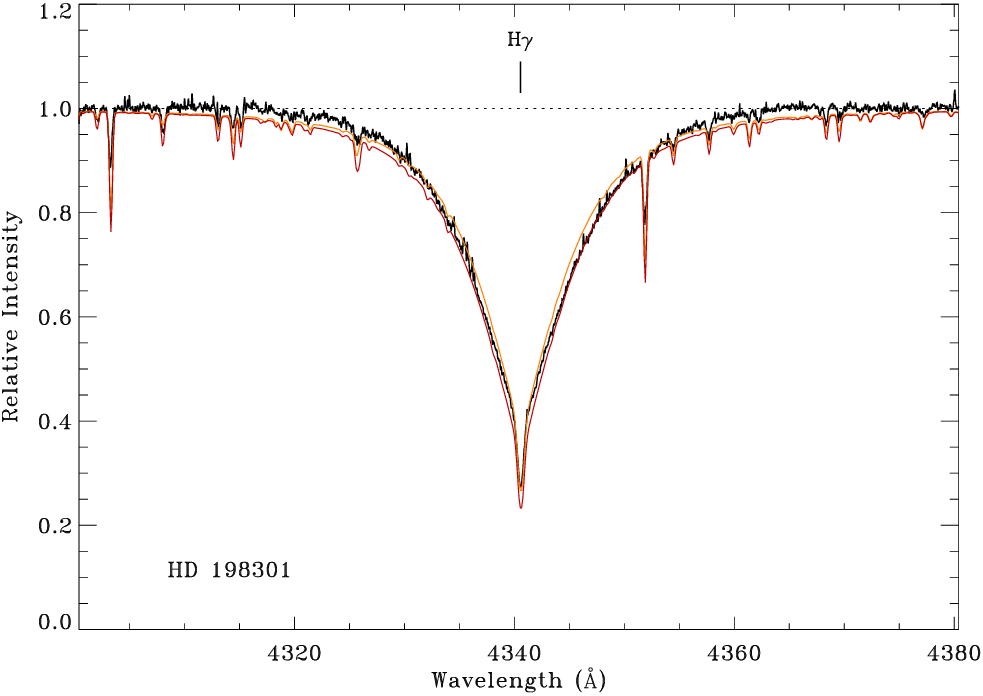}
\caption{Continuum normalized spectrum of HD 198301 in the vicinity of the stellar H$\gamma$ line. Model stellar spectra (colored curves) are shown superimposed onto the observed spectrum (black curve). The models correspond to $T_{\rm eff}=12,000$~K (red) and $13,000$~K (orange), with $\log g=3.5$ in both cases.}
\label{fig:hd198301_hgam}
\end{figure}

\subsubsection{HD~198301 --- a late B subgiant}
The procedure described in Appendix A yields a spectral type of B8 IV for HD~198301. We investigated the accuracy of this classification further using the high-resolution synthetic stellar spectra provided by \citet{m05}. The effective temperature of a B8 subgiant is expected to be $T_{\rm eff}\approx12,000$ K, while the surface gravity should be approximately $\log g\approx3.9$ \citep{sk82}. In Figure~\ref{fig:hd198301_hgam}, we show a comparison between the observed spectrum of HD~198301 in the vicinity of the stellar H$\gamma$ line and two synthetic stellar spectra. The model spectra correspond to $T_{\rm eff}=12,000$~K (red curve) and $13,000$~K (orange curve), with $\log g=3.5$ in both cases. (The models have been Doppler-shifted by 9 km~s$^{-1}$ to match the radial velocity of HD~198301.) Both models yield a relatively good fit to the data, although the higher temperature model has an H$\gamma$ line that is somewhat too narrow. (There is a slight mismatch in the wings of the line very far from the core. This is probably due to continuum placement errors. The Balmer lines generally span multiple echelle orders in our high-resolution spectra, making continuum placement challenging. However, this should not affect the widths of the lines closer to the core.)

\begin{figure}
\includegraphics[width=\columnwidth]{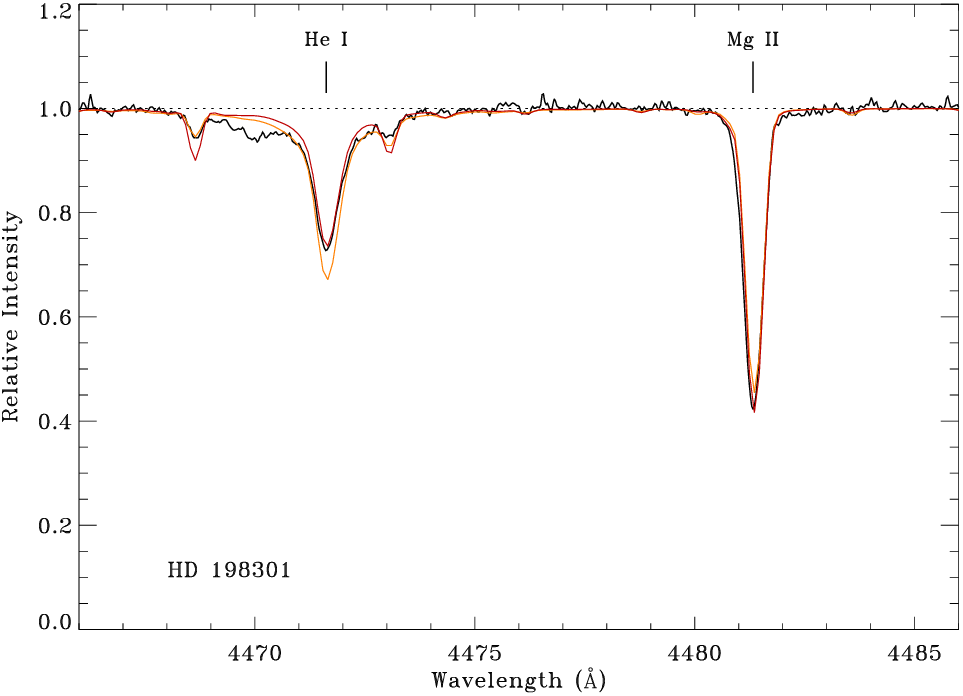}
\caption{Continuum normalized spectrum of HD 198301 in the vicinity of the stellar He~{\sc i} $\lambda4471$ and Mg~{\sc ii} $\lambda4481$ lines. Model stellar spectra (colored curves) are shown superimposed onto the observed spectrum (black curve). The models correspond to $T_{\rm eff}=12,000$~K (red) and $13,000$~K (orange), with $\log g=3.5$ in both cases.}
\label{fig:hd198301_hemg}
\end{figure}

In Figure~\ref{fig:hd198301_hemg}, we show the same two model stellar spectra compared with the observed spectrum of  HD~198301 in the vicinity of the stellar He~{\sc i}~$\lambda4471$ and Mg~{\sc ii}~$\lambda4481$ lines. The He~{\sc i}~$\lambda4471$/Mg~{\sc ii}~$\lambda4481$ line ratio is one of the main criteria used for temperature classification in late B-type stars \citep[e.g.,][]{gc09}. The good fit of the synthetic spectrum with $T_{\rm eff}=12,000$~K helps to firmly establish a temperature class of B8 for HD~198301. We note that a good match between the synthetic spectrum with $T_{\rm eff}=12,000$~K and $\log g=3.5$ and the observed spectrum of HD~198301 is found for other important stellar absorption lines, such as He~{\sc i}~$\lambda4026$, Si~{\sc ii}~$\lambda\lambda4128,4130$, and He~{\sc i}~$\lambda4921$.

\begin{figure}
\includegraphics[width=\columnwidth]{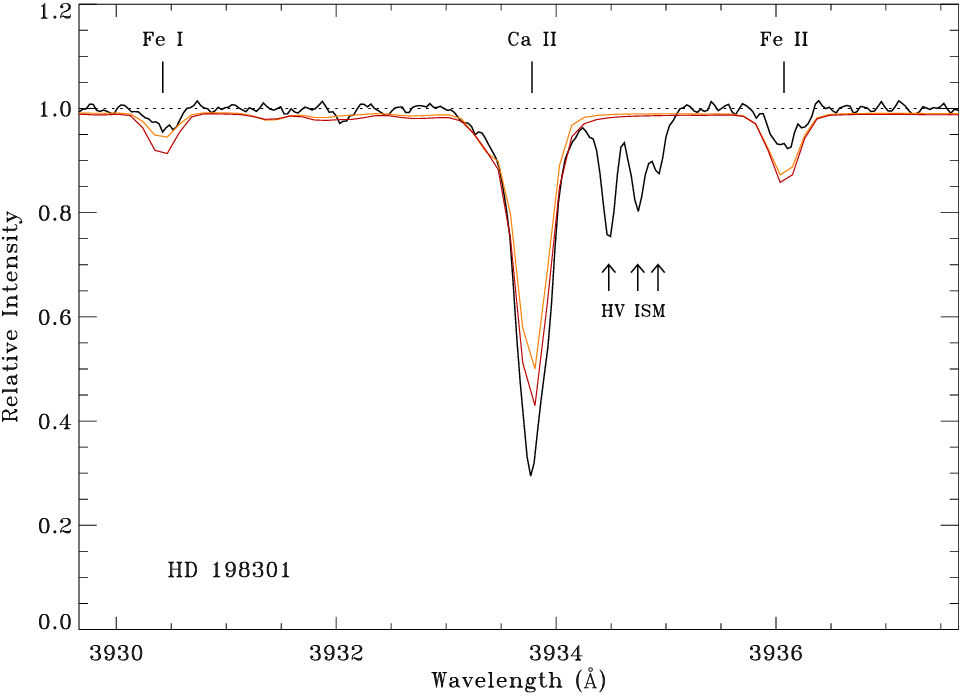}
\caption{Continuum normalized spectrum of HD 198301 in the vicinity of the Ca~{\sc ii} K line. Model stellar spectra (colored curves) are shown superimposed onto the observed spectrum (black curve). The models correspond to $T_{\rm eff}=12,000$~K (red) and $13,000$~K (orange), with $\log g=3.5$ in both cases. Three high-velocity interstellar Ca~{\sc ii} absorption features are identified.}
\label{fig:hd198301_cak}
\end{figure}

\begin{figure}
\includegraphics[width=\columnwidth]{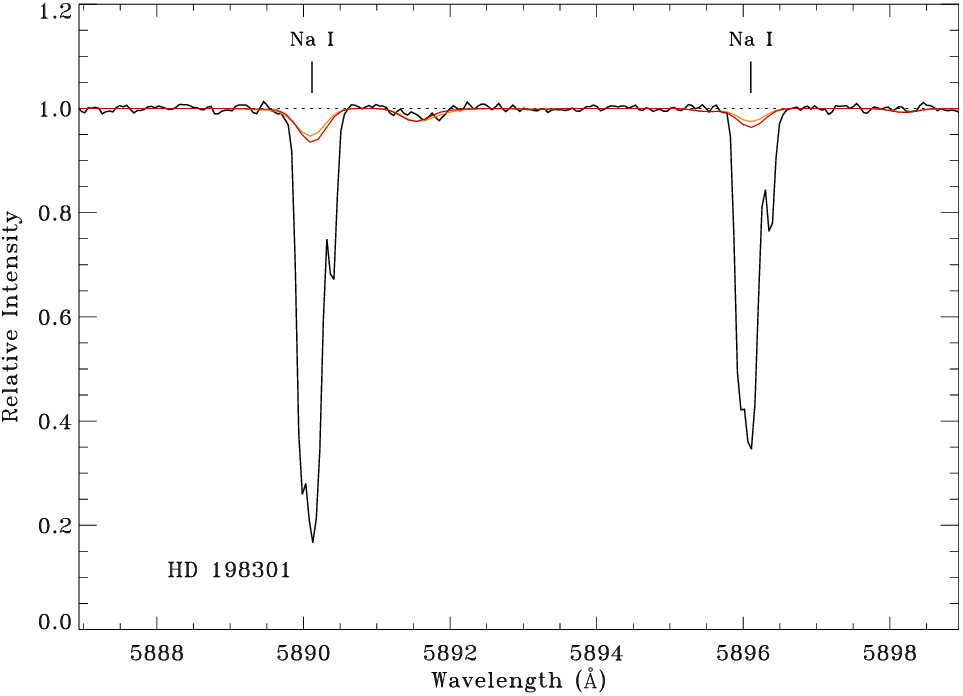}
\caption{Continuum normalized spectrum of HD 198301 in the vicinity of the Na~{\sc i} D lines. Model stellar spectra (colored curves) are shown superimposed onto the observed spectrum (black curve). The models correspond to $T_{\rm eff}=12,000$~K (red) and $13,000$~K (orange), with $\log g=3.5$ in both cases. Most of the Na~{\sc i} absorption is interstellar.}
\label{fig:hd198301_nad}
\end{figure}

Having established a best-fitting stellar model for HD~198301, we turned our attention to the Ca~{\sc ii}~K and Na~{\sc i}~D lines, where contributions from interstellar absorption lines are expected. Figure~\ref{fig:hd198301_cak} shows the Ca~{\sc ii}~K region. The observed Ca~{\sc ii}~K feature is deeper than either of the two synthetic models would predict, suggesting that some of this absorption may be due to interstellar Ca~{\sc ii} at low velocity. More striking, however, are the three narrow redshifted absorption features that fall between the strong Ca~{\sc ii}~K line and a nearby Fe~{\sc ii} line. The synthetic models have no absorption lines at these wavelengths. We therefore attribute these absorption features to high-velocity interstellar Ca~{\sc ii}~K components. The three components have LSR velocities of +62, +82, and +96 km~s$^{-1}$. Our interpretation of these absorption features is confirmed by the fact that the three components are detected at the same velocities in the nearby Ca~{\sc ii}~H line. (The components are somewhat harder to discern at Ca~{\sc ii}~H because they appear on the steeply decreasing edge of the stellar H$\epsilon$ line. Nevertheless, the reality of the components appears to be firmly established.)

The Na~{\sc i}~D region of the HD~198301 spectrum is shown in Figure~\ref{fig:hd198301_nad}. Here, it is obvious that most of the absorption in the Na~{\sc i}~D lines is from interstellar Na~{\sc i}. The Na~{\sc i} lines in both of the model stellar spectra are significantly weaker than the observed lines. Moreover, the observed Na~{\sc i} absorption profiles show a clear component structure of intrinsically narrow features, indicating an origin in cold, interstellar clouds. The high-velocity components seen in the Ca~{\sc ii} H and K lines are not detected in Na~{\sc i}. However, this is not surprising since high-velocity gas components associated with SNRs often have very low Na~{\sc i}/Ca~{\sc ii} ratios \citep[e.g.,][]{ds95,sw04,r20}, a consequence of the return of Ca$^+$ ions to the gas phase following the destruction of interstellar dust grains in SNR shocks.

Before proceeding with the analysis of the interstellar Ca~{\sc ii}~H and K profiles and the Na~{\sc i}~D profiles toward HD~198301, the observed spectra for those regions were divided by the best-fitting model spectrum (i.e., the model with $T_{\rm eff}=12,000$~K and $\log g=3.5$) in order to remove the contributions from stellar absorption lines. This procedure had little effect on the interstellar Na~{\sc i}~D profiles or on the high-velocity Ca~{\sc ii} components. However, the low-velocity interstellar Ca~{\sc ii} absorption toward HD~198301 is severely impacted by stellar absorption and the removal of this contaminating absorption is model-dependent. While we have chosen the best stellar model available, there remains some uncertainty about the strength of the interstellar Ca~{\sc ii} line at low velocity.

\subsubsection{HD~335334 --- a double-line spectroscopic binary}
Initially, our high-resolution spectrum of HD~335334 was difficult to interpret. There appeared to be many more narrow stellar absorption lines than would be expected for a late B or early A dwarf or subgiant. Upon closer examination, it became clear that each of the major stellar absorption lines had a corresponding line at a fixed velocity from the primary line. We thus suspected that the star was a double-line spectroscopic binary. Fortunately, this star had been observed on two consecutive nights during the six night observing run. Two 30 minute exposures of HD~335334 were obtained on 2022 Sep.~4, and six 30 minute exposures were obtained on Sep.~5. (UT dates are quoted.) We therefore created summed spectra for each night separately to see if there was any shift in the stellar absorption lines from one night to the next.

\begin{figure}
\includegraphics[width=\columnwidth]{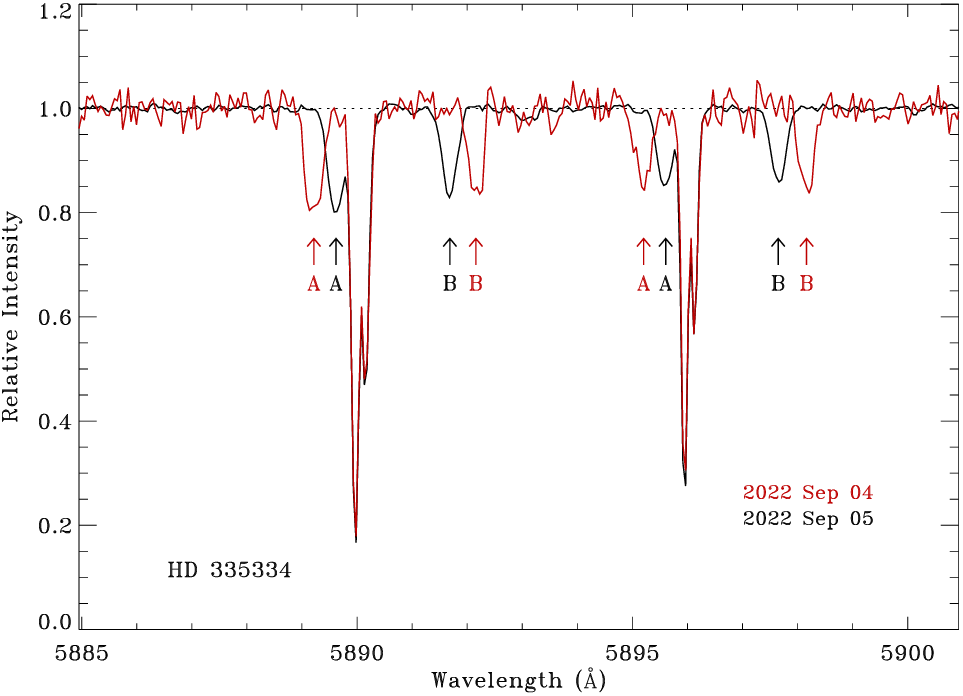}
\caption{Continuum normalized spectra of HD 335334 in the vicinity of the Na~{\sc i} D lines from observations obtained on two consecutive nights: 2022 Sep. 4 (red) and Sep. 5 (black). The stellar Na~{\sc i} D lines of the primary component (``A'') and the secondary component (``B'') of the spectroscopic binary can be seen to shift from one night to the next, while the interstellar Na~{\sc i}~D lines are stationary.}
\label{fig:hd335334_nad}
\end{figure}

In Figure~\ref{fig:hd335334_nad}, we show the separate summed spectra of HD~335334 from the two consecutive nights for the spectral region containing the Na~{\sc i}~D lines. The spectrum from Sep.~4 (plotted with a red line) has a lower S/N ratio due to the lower total exposure time. Nevertheless, it is clear that the stellar Na~{\sc i} lines (labelled ``A'' and ``B'' in the figure) shifted significantly between the two observations, while the two narrow interstellar Na~{\sc i} components remained stationary. In the spectrum acquired on Sep.~4, the radial velocities of the stellar Na~{\sc i} lines are $-$37 km~s$^{-1}$ for component A and +113 km~s$^{-1}$ for component B. In the Sep.~5 spectrum, the velocities are $-$17 km~s$^{-1}$ for component A and +88 km~s$^{-1}$ for component B. Thus, the velocity shift for component A is $\sim$20 km~s$^{-1}$ (to the red), while the shift for component B is $\sim$25 km~s$^{-1}$ (to the blue). The slightly smaller shift for component A indicates that this star has a somewhat higher mass (hence the designation as component ``A'').

\begin{figure}
\includegraphics[width=\columnwidth]{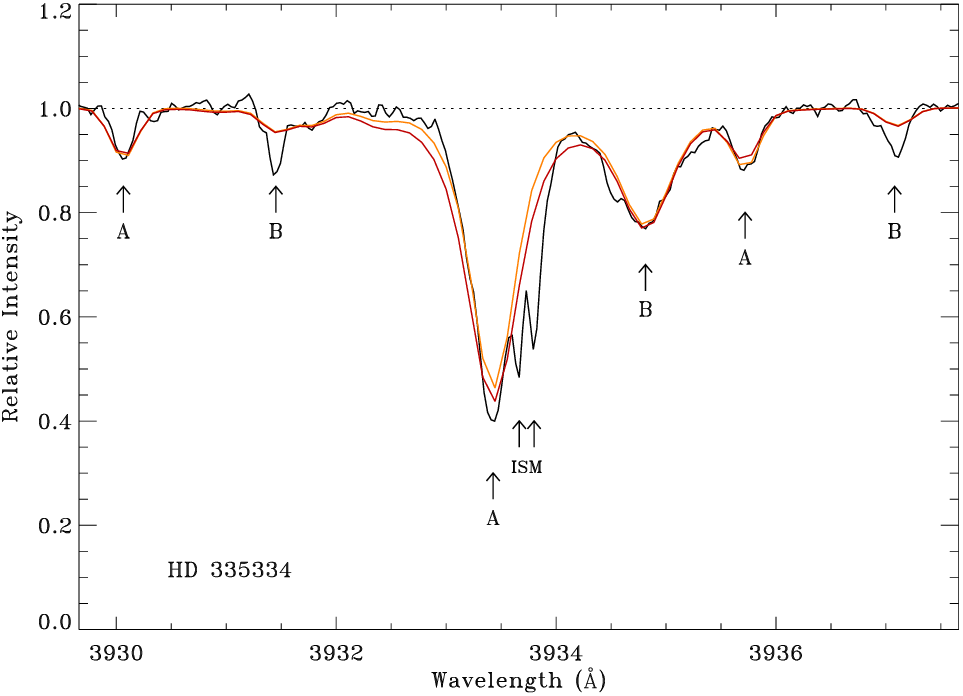}
\caption{Continuum normalized spectrum of HD 335334 in the vicinity of the Ca~{\sc ii} K line from our observations obtained on 2022 Sep. 5. Model stellar spectra (colored curves) are superimposed onto the observed spectrum (black curve). The models show composite stellar spectra, with $T_{\rm eff}=10,500$~K for both the primary component (``A'') and the secondary component (``B''). The red curve adopts $\log g=4.0$ for both the primary and the secondary, while the orange curve adopts $\log g=3.5$ for the primary and $\log g=4.0$ for the secondary. Two low-velocity interstellar Ca~{\sc ii} components are identified. These are the same two interstellar components that can be seen in the Na~{\sc i} D lines in Figure~\ref{fig:hd335334_nad}.}
\label{fig:hd335334_cak}
\end{figure}

The Ca~{\sc ii}~K region of the Sep.~5 spectrum of HD~335334 is shown in Figure~\ref{fig:hd335334_cak}. Two synthetic model spectra are also shown in the plot. The model spectra were constructed from the high-resolution synthetic spectra provided by \citet{m05}. However, in this case, we created composite spectra to compare with the observations. The procedure described in Appendix A yields a spectral type of B9 V for HD~335334.\footnote{The procedure we use for classifying our stars is based on an analysis of the spectra after smoothing the data to a resolution of 1.8~\AA{} (see Appendix A). At such a low resolution, the two components of the spectroscopic binary cannot be discerned. Thus, the spectral type we derive for HD~335334 refers to the composite system only.} A B9 V star is expected to have an effective temperature of $T_{\rm eff}\approx10,500$ K and a surface gravity of $\log g\approx4.0$ \citep{sk82}. Thus, we created a composite spectrum from two synthetic model spectra, both with $T_{\rm eff}=10,500$ K and $\log g=4.0$. We Doppler-shifted the model spectra in accordance with the observed radial velocities of the two components. We then divided the flux of the model for component B by 2.5 and summed the two model spectra together to produce a composite spectrum. The normalized version of this composite spectrum is shown by the red curve in Figure~\ref{fig:hd335334_cak}. The factor of 2.5 was determined by trial-and-error to match the observed ratio of the Ca~{\sc ii}~K lines from components A and B (see Figure~\ref{fig:hd335334_cak}). This same ratio between A and B is seen in several other prominent absorption lines in the spectrum of HD~335334 and may indicate that the luminosity of component A is $\sim$2.5 times larger than that of component B. Given the prospect of a higher luminosity for component A, we created another composite model spectrum with all of the same parameters as the first model except the surface gravity of the primary component was changed to $\log g=3.5$. This alternative composite spectrum is shown by the orange curve in Figure~\ref{fig:hd335334_cak}.

The purpose of creating the composite spectra discussed above was not to attempt to match the observed spectrum of HD~335334 perfectly. The main purpose was to correctly identify stellar absorption features particularly in the vicinity of the Ca~{\sc ii}~H and K lines, so that the interstellar Ca~{\sc ii} absorption features could be properly identified and analyzed. As can be seen in Figure~\ref{fig:hd335334_cak}, much of the absorption in the Ca~{\sc ii}~K region of the HD~335334 spectrum arises from stellar lines. The strongest feature is the Ca~{\sc ii}~K line from the primary star. The secondary Ca~{\sc ii}~K line is the other broad absorption feature redshifted by 105 km~s$^{-1}$ relative to the primary. Two other pairs of stellar lines can be seen in the figure. These are the Fe~{\sc i}~$\lambda3930$ and Fe~{\sc ii}~$\lambda3935$ lines that are also seen in the spectrum of HD~198301 (Figure~\ref{fig:hd198301_cak}). Two narrow interstellar Ca~{\sc ii} components can be seen on the red side of the primary Ca~{\sc ii}~K line. These are the same two low-velocity interstellar components that can be seen in the Na~{\sc i}~D lines toward HD~335334 (Figure~\ref{fig:hd335334_nad}). After accounting for stellar absorption lines, we find no evidence of high-velocity interstellar Na~{\sc i} or Ca~{\sc ii} components toward HD~335334.

\citet{f18b} claimed to detect high-velocity interstellar Na~{\sc i} and Ca~{\sc ii} components toward HD~335334 (a star that those authors refer to as ``Star X''). They reported finding a blueshifted component in Na~{\sc i} and Ca~{\sc ii} at approximately $-$60 km~s$^{-1}$ and a redshifted component near +90 km~s$^{-1}$. However, considering the analysis presented above, it is now clear that the ``blue'' and ``red'' components discussed by \citet{f18b} are not high-velocity interstellar components. They are the stellar Na~{\sc i} and Ca~{\sc ii} components from the primary and secondary stars that constitute the spectroscopic binary. Note that the ``red'' component discussed by \citet{f18b} is the primary and the ``blue'' component is the secondary. This can be seen in their Figure~3, which shows that the Ca~{\sc ii} absorption is much stronger in the red component compared to the blue component. Clearly, \citet{f18b} observed HD~335334 at a different orbital phase compared to our observations.

Before proceeding with the analysis of the interstellar Na~{\sc i} and Ca~{\sc ii} lines toward HD~335334, we removed the surrounding stellar absorption features. In this case, we did not use the model stellar spectra to divide out the stellar absorption (because the models are not a perfect match to the observed spectrum). Instead, we manually removed the obvious stellar absorption features. For the Na~{\sc i} lines, this was not a problem because the interstellar Na~{\sc i} lines are much stronger than the stellar lines. For the Ca~{\sc ii} H and K lines, where the interstellar features are blended with stellar absorption from the primary, we did our best to deblend the various features. Nevertheless, there remains some uncertainty about the strength of the low-velocity interstellar Ca~{\sc ii} components toward HD~335334.

\subsection{Profile fitting}
After removing any contaminating stellar absorption from the interstellar absorption profiles toward our targets, the interstellar lines were analyzed using a multi-component Voigt profile fitting routine known as ISMOD \citep{s08}. The profile synthesis code derives the best-fitting column density, velocity, and Doppler $b$-value of each individual interstellar component through a simple root mean square minimizing procedure. An additional constraint for the Na~{\sc i} and Ca~{\sc ii} analyses is that the two lines of the doublet (Na~{\sc i}~D$_1$ and D$_2$ and Ca~{\sc ii}~H and K) were fit simultaneously so that a single set of component parameters is obtained for each doublet. While our spectra have moderately high resolution ($\sim$4.5 km~s$^{-1}$), we are likely still not resolving the individual interstellar absorption components. High and ultra-high resolution studies of interstellar lines have revealed that the intrinsic widths of Ca~{\sc ii} components are typically in the range 1 to 3 km~s$^{-1}$ \citep[e.g.,][]{w96,p05}, while Na~{\sc i} components often have intrinsic $b$-values that are $\sim$1 km~s$^{-1}$ or less \citep[e.g.,][]{w94}. A simultaneous fit to both members of the Na~{\sc i} and Ca~{\sc ii} doublets can thus help to mitigate the effect on the derived column densities of unresolved saturation in the line profiles (particularly for the Na~{\sc i} lines).

\begin{figure}
\includegraphics[width=\columnwidth]{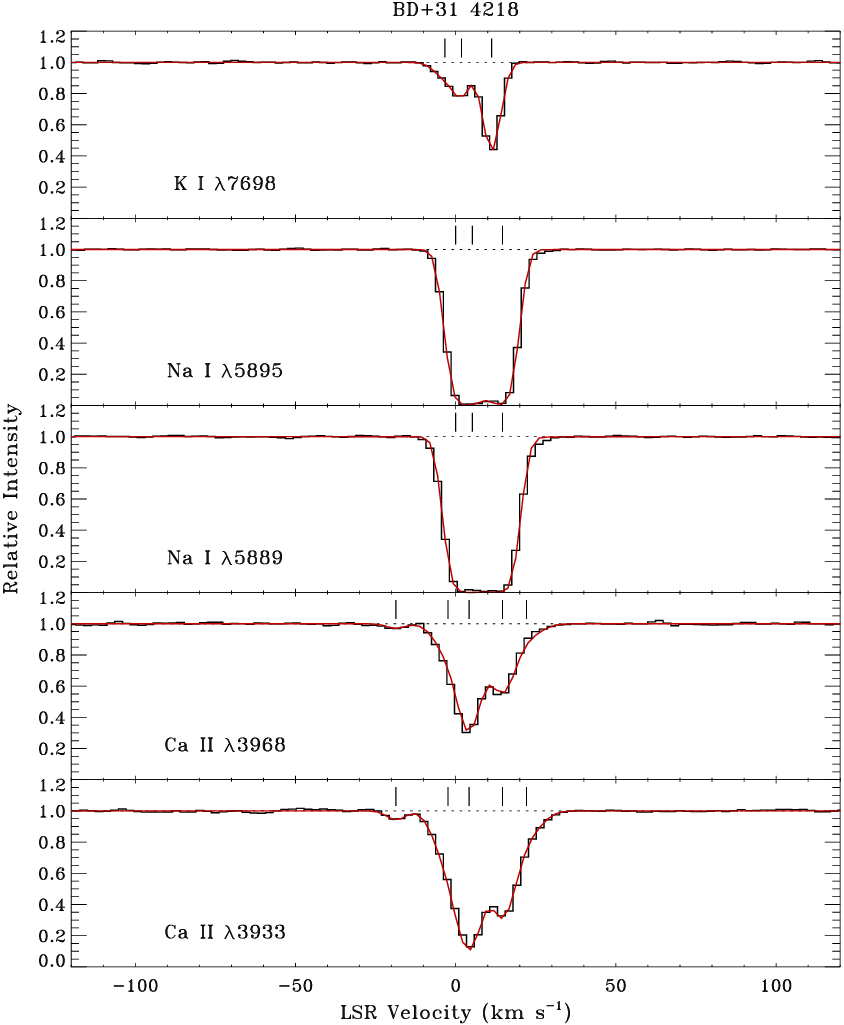}
\caption{Simultaneous profile synthesis fits to the interstellar Na~{\sc i} D$_1$ and D$_2$ lines and the Ca~{\sc ii} H and K lines toward BD+31~4218. An independent fit to the K~{\sc i} $\lambda7698$ line is also shown. Synthetic absorption profiles (red curves) are shown superimposed onto the observed spectra (black histograms). Tick marks give the positions of the individual components included in the fits.}
\label{fig:bd314218_nacak}
\end{figure}

For most of our target sight lines, the only strong interstellar absorption features detected are the Na~{\sc i}~D and Ca~{\sc ii}~H and K lines. In some cases, a weak K~{\sc i}~$\lambda7698$ feature is also detected. This is consistent with the low reddening seen toward most of our stars (Table~\ref{tab:targets}). With the exception of BD+31~4218, the values of $E$($B$$-$$V$) for our sight lines are $\sim$0.1 or less, consistent with the value of 0.08 typically quoted for the Cygnus Loop \citep[e.g.,][]{r81,f82}. The reddening toward BD+31~4218, however, is $\sim$0.3. This star has the largest distance among the stars in our sample ($\sim$1100 pc) and is located somewhat beyond the bright optical emission features to the northwest of the Cygnus Loop (see Figure~\ref{fig:cygnus_dss2}). \citet{f18a,f18b} demonstrated that the region of the sky immediately beyond the western limb of the Cygnus Loop is characterized by a sharp increase in dust extinction, consistent with our finding a higher value of $E$($B$$-$$V$) toward BD+31~4218. Several other atomic and molecular absorption lines are detected toward BD+31~4218, including Li~{\sc i}~$\lambda6707$, Ca~{\sc i}~$\lambda4226$, CH~$\lambda4300$, CH$^+$~$\lambda4232$, and CN~$\lambda3874$. These lines likely arise from material associated with the western molecular cloud that \citet{f18b} suggested may be interacting with the Cygnus Loop.

Figure~\ref{fig:bd314218_nacak} presents our simultaneous profile synthesis fits to the Na~{\sc i} and Ca~{\sc ii} lines toward BD+31~4218. Because the Na~{\sc i}~D lines are heavily saturated in this direction, we first fit the K~{\sc i}~$\lambda7698$ line independently. (This fit is shown in the top panel of Figure~\ref{fig:bd314218_nacak}.) We then held the fractional column densities and relative velocities of the three K~{\sc i} components fixed in our simultaneous fit to the Na~{\sc i}~D lines. The resulting $N$(Na~{\sc i})/$N$(K~{\sc i}) ratio of $\sim$80 for the line of sight is consistent with the ratios that characterize typical diffuse molecular cloud sight lines \citep[e.g.,][]{wh01}.

\begin{figure}
\includegraphics[width=\columnwidth]{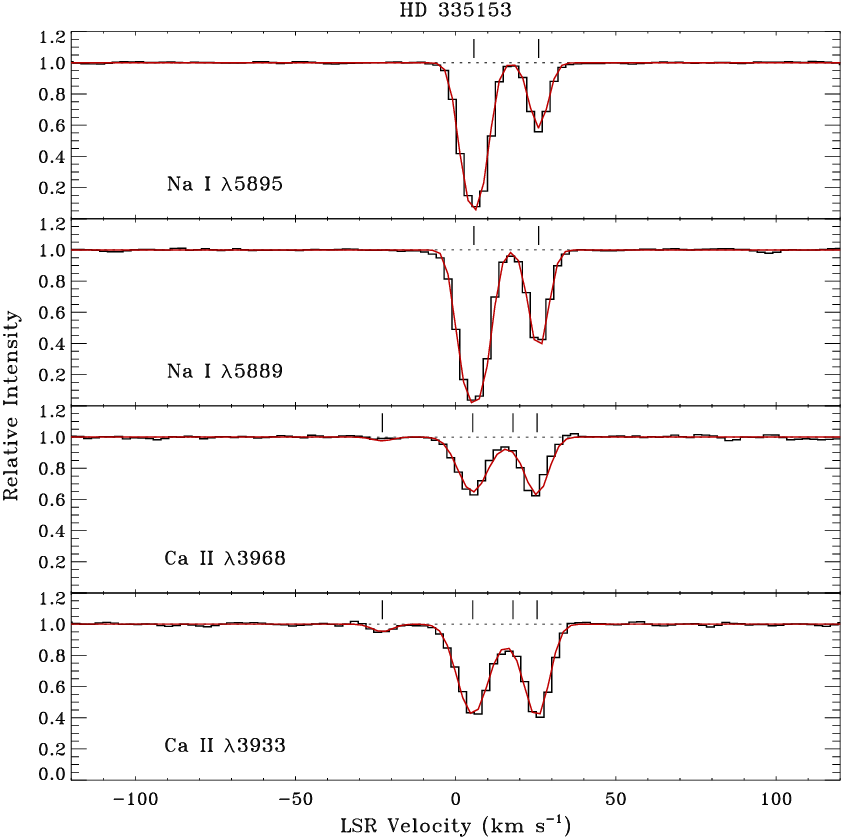}
\caption{Simultaneous profile synthesis fits to the interstellar Na~{\sc i} D$_1$ and D$_2$ lines and the Ca~{\sc ii} H and K lines toward HD~335153. Synthetic absorption profiles (red curves) are shown superimposed onto the observed spectra (black histograms). Tick marks give the positions of the individual components included in the fits.}
\label{fig:hd335153_naca}
\end{figure}

\begin{figure}
\includegraphics[width=\columnwidth]{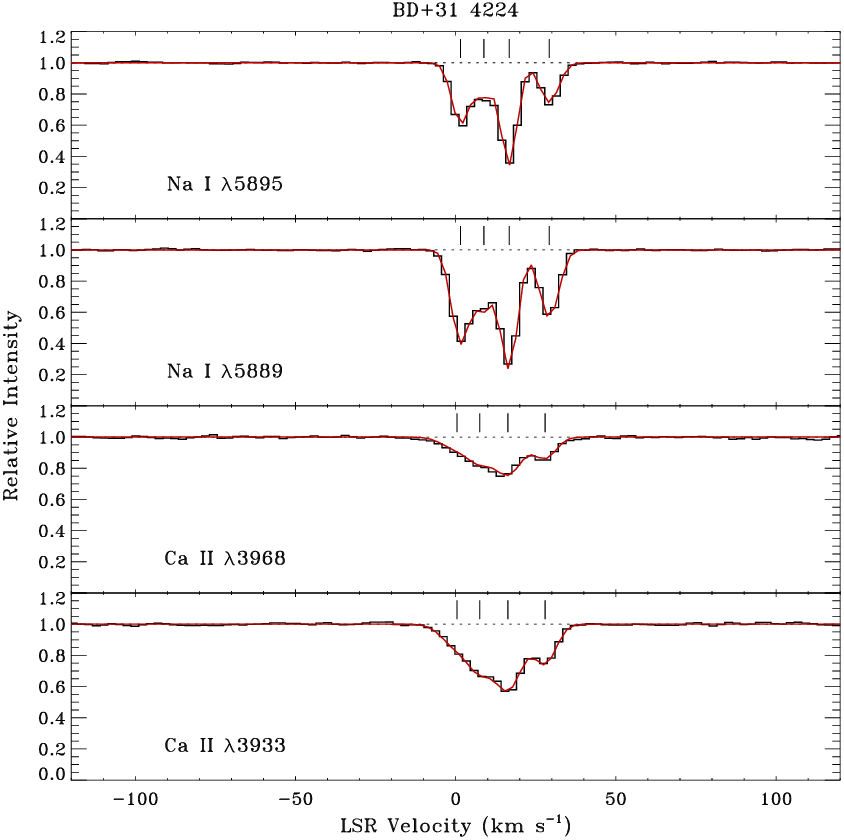}
\caption{Simultaneous profile synthesis fits to the interstellar Na~{\sc i} D$_1$ and D$_2$ lines and the Ca~{\sc ii} H and K lines toward BD+31~4224. Synthetic absorption profiles (red curves) are shown superimposed onto the observed spectra (black histograms). Tick marks give the positions of the individual components included in the fits.}
\label{fig:bd314224_naca}
\end{figure}

\begin{figure}
\includegraphics[width=\columnwidth]{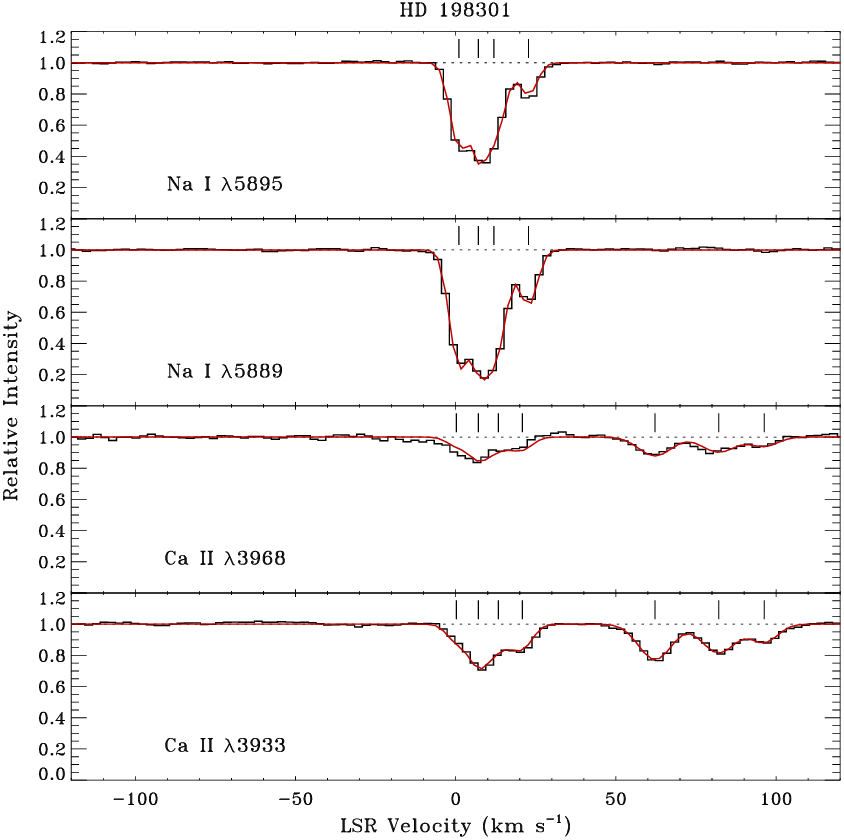}
\caption{Simultaneous profile synthesis fits to the interstellar Na~{\sc i} D$_1$ and D$_2$ lines and the Ca~{\sc ii} H and K lines toward HD~198301. Synthetic absorption profiles (red curves) are shown superimposed onto the observed spectra (black histograms). Tick marks give the positions of the individual components included in the fits. Note the high-velocity Ca~{\sc ii} components at +62, +82, and +96 km~s$^{-1}$.}
\label{fig:hd198301_naca}
\end{figure}

\begin{figure}
\includegraphics[width=\columnwidth]{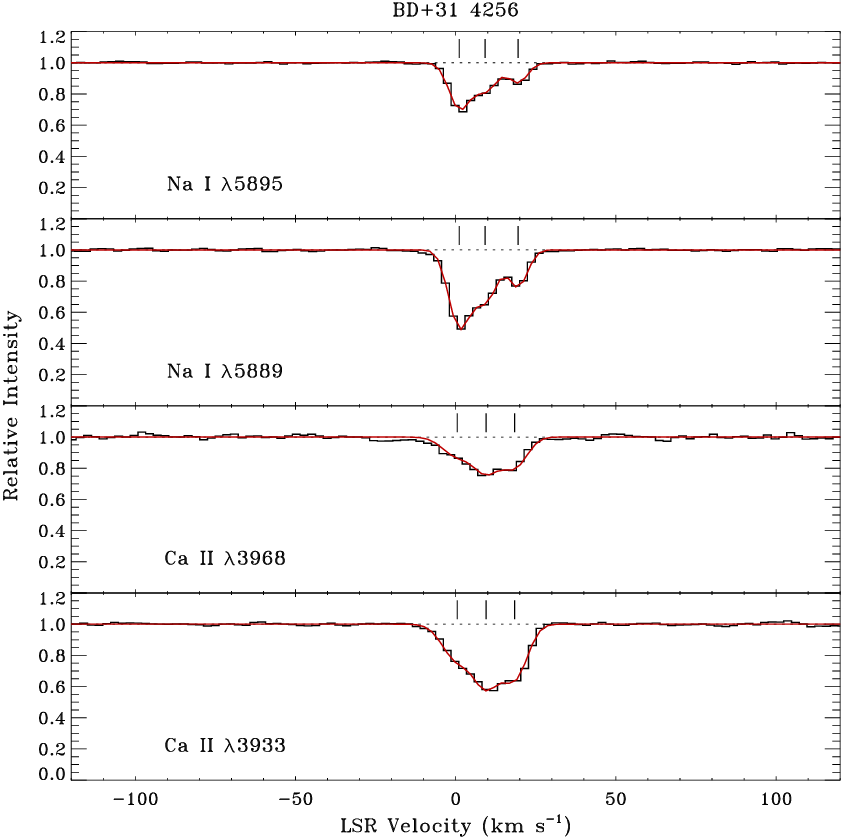}
\caption{Simultaneous profile synthesis fits to the interstellar Na~{\sc i} D$_1$ and D$_2$ lines and the Ca~{\sc ii} H and K lines toward BD+31~4256. Synthetic absorption profiles (red curves) are shown superimposed onto the observed spectra (black histograms). Tick marks give the positions of the individual components included in the fits.}
\label{fig:bd314256_naca}
\end{figure}

\begin{figure}
\includegraphics[width=\columnwidth]{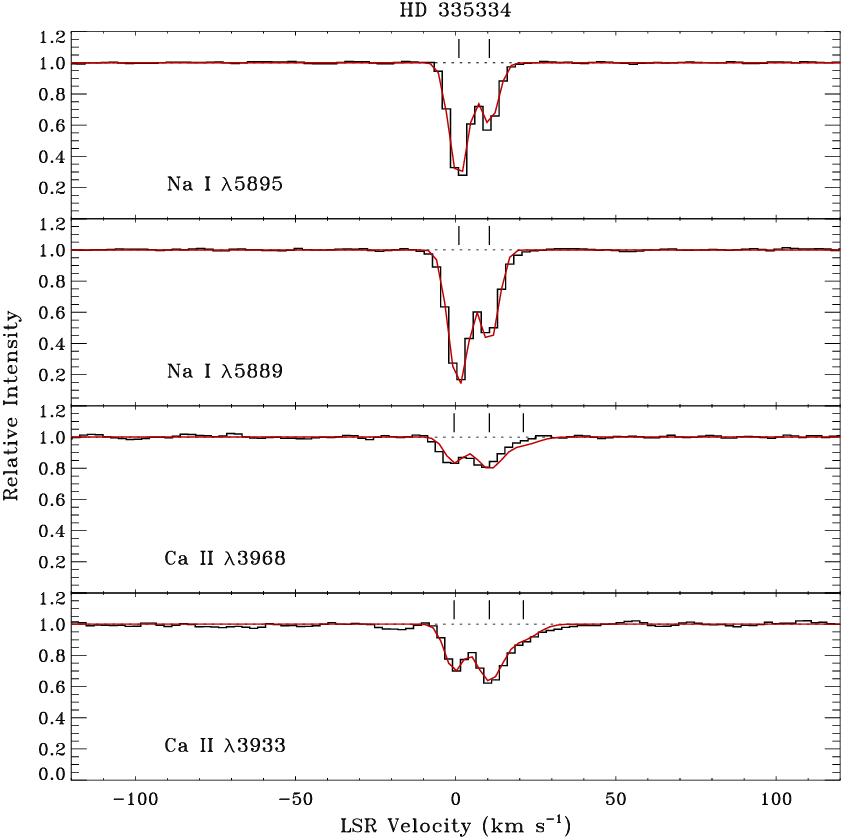}
\caption{Simultaneous profile synthesis fits to the interstellar Na~{\sc i} D$_1$ and D$_2$ lines and the Ca~{\sc ii} H and K lines toward HD~335334. Synthetic absorption profiles (red curves) are shown superimposed onto the observed spectra (black histograms). Tick marks give the positions of the individual components included in the fits.}
\label{fig:hd335334_naca}
\end{figure}

Our simultaneous profile synthesis fits to the interstellar Na~{\sc i}~D and Ca~{\sc ii}~H and K lines toward the other five targets are presented in Figures~\ref{fig:hd335153_naca} through \ref{fig:hd335334_naca}. In general, these sight lines show much weaker Na~{\sc i} and Ca~{\sc ii} absorption features and the process of fitting the absorption profiles did not present any unusual complications. We did strive to include Na~{\sc i} and Ca~{\sc ii} components at similar velocities for a given sight line so that the $N$(Na~{\sc i})/$N$(Ca~{\sc ii}) ratios of the different components could be analyzed in a consistent manner. Again, at a velocity resolution of $\sim$4.5 km~s$^{-1}$, we are likely not fully resolving the detailed interstellar component structure toward our targets. Nevertheless, given the general weakness of the absorption lines, and our procedure of simultaneously fitting both lines of the Na~{\sc i} and Ca~{\sc ii} doublets, the column densities we derive should not be significantly impacted.

Table~\ref{tab:totals} presents the total equivalent widths and column densities of the interstellar atomic and molecular lines detected toward our sample of stars. The equivalent width errors reflect uncertainties due to noise in the continuum as well as errors due to continuum placement. The column density errors include an additional term related to the level of saturation in the absorption line. As anticipated, the Na~{\sc i} column density toward BD+31~4218 is more than an order of magnitude larger than toward the other targets. Comparisons of the total (line-of-sight) column densities of the atomic and molecular species observed toward BD+31~4218 (e.g., Na~{\sc i} vs.~K~{\sc i}, Li~{\sc i} vs.~Na~{\sc i} and K~{\sc i}, and CH vs.~Na~{\sc i} and K~{\sc i}) all show very good agreement with typical Galactic relationships \citep{wh01,w06}. The $N$(Na~{\sc i}) and $N$(K~{\sc i}) values then imply a total hydrogen column density of $\log N({\rm H}_{\rm tot})\sim21.3$ toward BD+31~4218 \citep{wh01}. Likewise, the column densities of Na~{\sc i}, K~{\sc i}, and CH imply a molecular hydrogen column density of $\log N({\rm H}_2)\sim20.8$ and a molecular fraction of $f({\rm H}_2)=2N({\rm H}_2)/N({\rm H}_{\rm tot})\sim0.6$ \citep{wh01,s08}. For the other five sight lines, the Na~{\sc i} and K~{\sc i} column densities imply total hydrogen column densities in the range $\log N({\rm H}_{\rm tot})\sim20.4$--$20.8$ and molecular hydrogen fractions of $f({\rm H}_2)\sim0.1$.

\begin{table}
\centering
\caption{Total equivalent widths (in m\AA{}) and column densities (in cm$^{-2}$) of the atomic and molecular species observed toward the program stars. The three CN lines listed toward BD+31~4218 are (in order of increasing wavelength) the $R$(1), $R$(0), and $P$(1) lines of the $B$$-$$X$ (0,~0) band.}
\label{tab:totals}
\begin{tabular}{lcccc}
\hline
Star & Species & $\lambda$ & $W_{\lambda}$ & $\log N$ \\
 & & (\AA{}) & (m\AA{}) & \\
\hline
BD+31~4218 & Li~{\sc i} & 6707.826 & $1.9\pm0.6$ & $9.81\pm0.12$ \\
 & Na~{\sc i} & 5889.951 & $491.3\pm0.6$ & $14.02\pm0.07$ \\
 & & 5895.924 & $473.8\pm0.7$ & $14.02\pm0.07$ \\
 & K~{\sc i} & 7698.965 & $155.7\pm1.0$ & $12.11\pm0.03$ \\
 & Ca~{\sc i} & 4226.728 & $8.9\pm0.8$ & $10.51\pm0.04$ \\
 & Ca~{\sc ii} & 3933.661 & $251.5\pm1.2$ & $12.74\pm0.03$ \\
 & & 3968.467 & $170.8\pm1.4$ & $12.74\pm0.02$ \\
 & CH & 4300.313 & $19.8\pm0.5$ & $13.42\pm0.02$ \\
 & CH$^+$ & 4232.548 & $5.1\pm0.7$ & $12.78\pm0.05$ \\
 & CN & 3873.994 & $5.9\pm0.5$ & $12.33\pm0.04$ \\
 & & 3874.602 & $18.4\pm0.6$ & $12.67\pm0.02$ \\
 & & 3875.758 & $4.1\pm0.8$ & $12.33\pm0.07$ \\
HD~335153 & Na~{\sc i} & 5889.951 & $311.5\pm1.0$ & $12.86\pm0.07$ \\
 & & 5895.924 & $261.0\pm1.0$ & $12.86\pm0.06$ \\
 & K~{\sc i} & 7698.965 & $18.6\pm0.9$ & $11.04\pm0.02$ \\
 & Ca~{\sc ii} & 3933.661 & $181.5\pm1.5$ & $12.46\pm0.02$ \\
 & & 3968.467 & $98.9\pm1.6$ & $12.46\pm0.01$ \\
BD+31~4224 & Na~{\sc i} & 5889.951 & $292.8\pm0.9$ & $12.55\pm0.05$ \\
 & & 5895.924 & $209.4\pm0.9$ & $12.55\pm0.04$ \\
 & K~{\sc i} & 7698.965 & $5.5\pm0.7$ & $10.50\pm0.05$ \\
 & Ca~{\sc ii} & 3933.661 & $134.6\pm1.2$ & $12.28\pm0.01$ \\
 & & 3968.467 & $79.5\pm1.8$ & $12.28\pm0.01$ \\
HD~198301 & Na~{\sc i} & 5889.951 & $325.8\pm1.0$ & $12.59\pm0.03$ \\
 & & 5895.924 & $236.4\pm1.1$ & $12.59\pm0.02$ \\
 & Ca~{\sc ii} & 3933.661 & $155.4\pm2.0$ & $12.30\pm0.01$ \\
 & & 3968.467 & $83.6\pm3.9$ & $12.30\pm0.02$ \\
BD+31~4256 & Na~{\sc i} & 5889.951 & $169.8\pm1.2$ & $12.05\pm0.02$ \\
 & & 5895.924 & $98.9\pm1.3$ & $12.05\pm0.01$ \\
 & Ca~{\sc ii} & 3933.661 & $124.8\pm1.6$ & $12.26\pm0.01$ \\
 & & 3968.467 & $71.5\pm2.5$ & $12.26\pm0.02$ \\
HD~335334 & Na~{\sc i} & 5889.951 & $214.7\pm0.7$ & $12.50\pm0.06$ \\
 & & 5895.924 & $165.7\pm0.8$ & $12.50\pm0.04$ \\
 & K~{\sc i} & 7698.965 & $15.7\pm1.1$ & $10.97\pm0.03$ \\
 & Ca~{\sc ii} & 3933.661 & $93.5\pm2.3$ & $12.10\pm0.02$ \\
 & & 3968.467 & $47.9\pm1.7$ & $12.10\pm0.02$ \\
\hline
\end{tabular}
\end{table}

\begin{table*}
\centering
\caption{Velocities (in km~s$^{-1}$), column densities (in cm$^{-2}$), and $b$-values (in km~s$^{-1}$) of the individual Ca~{\sc ii} and Na~{\sc i} absorption components discerned through profile fitting. Upper limits on $N$(Na~{\sc i}) are provided in cases where Ca~{\sc ii} is detected but Na~{\sc i} is not. The last column gives the $N$(Na~{\sc i})/$N$(Ca~{\sc ii}) ratio (or upper limit on the ratio).}
\label{tab:components}
\begin{tabular}{lccccccc}
\hline
Star & $v_{\rm LSR}$(Ca~{\sc ii}) & $\log N$(Ca~{\sc ii}) & $b$(Ca~{\sc ii}) & $v_{\rm LSR}$(Na~{\sc i}) & $\log N$(Na~{\sc i}) & $b$(Na~{\sc i}) & $N$(Na~{\sc i})/$N$(Ca~{\sc ii}) \\
\hline
BD+31~4218 & $-$18.7 & $10.81\pm0.04$ & 2.7 & \ldots & $<$9.8 & \ldots & $<$0.10 \\
 & $-$2.4 & $11.78\pm0.01$ & 4.8 & +0.0 & $12.98\pm0.06$ & 2.2 & $16.0\pm2.5$ \\
 & +4.2 & $12.48\pm0.03$ & 3.7 & +5.2 & $13.26\pm0.07$ & 2.4 & $6.1\pm1.1$ \\
 & +14.6 & $12.20\pm0.02$ & 4.3 & +14.6 & $13.88\pm0.07$ & 2.4 & $47.9\pm8.6$ \\
 & +22.1 & $11.44\pm0.02$ & 5.5 & \ldots & $<$9.8 & \ldots & $<$0.02 \\
HD~335153 & $-$22.9 & $10.77\pm0.05$ & 3.3 & \ldots & $<$10.1 & \ldots & $<$0.24 \\
 & +5.3 & $12.15\pm0.02$ & 5.3 & +5.7 & $12.81\pm0.05$ & 3.4 & $4.55\pm0.63$ \\
 & +17.9 & $11.27\pm0.02$ & 5.4 & \ldots & $<$10.1 & \ldots & $<$0.07 \\
 & +25.4 & $12.08\pm0.02$ & 3.9 & +25.9 & $11.90\pm0.02$ & 2.6 & $0.67\pm0.04$ \\
BD+31~4224 & +0.4 & $11.42\pm0.02$ & 5.5 & +1.5 & $11.84\pm0.02$ & 2.3 & $2.62\pm0.17$ \\
 & +7.5 & $11.62\pm0.01$ & 4.1 & +8.8 & $11.55\pm0.01$ & 2.8 & $0.85\pm0.03$ \\
 & +16.3 & $11.94\pm0.01$ & 5.0 & +16.7 & $12.32\pm0.05$ & 1.6 & $2.42\pm0.29$ \\
 & +27.9 & $11.58\pm0.01$ & 3.7 & +29.2 & $11.64\pm0.01$ & 2.9 & $1.15\pm0.05$ \\
HD~198301 & +0.2 & $11.10\pm0.02$ & 3.0 & +1.0 & $12.07\pm0.03$ & 2.3 & $9.31\pm0.85$ \\
 & +7.1 & $11.55\pm0.01$ & 3.5 & +7.1 & $12.11\pm0.05$ & 1.2 & $3.68\pm0.43$ \\
 & +13.3 & $11.36\pm0.02$ & 4.6 & +11.9 & $12.04\pm0.02$ & 3.4 & $4.78\pm0.33$ \\
 & +20.8 & $11.29\pm0.02$ & 3.4 & +22.7 & $11.49\pm0.01$ & 2.4 & $1.61\pm0.08$ \\
 & +62.2 & $11.69\pm0.01$ & 6.4 & \ldots & $<$10.0 & \ldots & $<$0.02 \\
 & +82.1 & $11.60\pm0.01$ & 6.5 & \ldots & $<$10.0 & \ldots & $<$0.02 \\
 & +96.3 & $11.30\pm0.02$ & 5.1 & \ldots & $<$10.0 & \ldots & $<$0.05 \\
BD+31~4256 & +0.5 & $11.65\pm0.02$ & 5.5 & +1.1 & $11.74\pm0.01$ & 3.3 & $1.22\pm0.06$ \\
 & +9.5 & $11.87\pm0.02$ & 4.6 & +9.2 & $11.57\pm0.01$ & 4.2 & $0.51\pm0.02$ \\
 & +18.4 & $11.79\pm0.02$ & 4.5 & +19.5 & $11.29\pm0.01$ & 2.8 & $0.31\pm0.01$ \\
HD~335334 & $-$0.5 & $11.59\pm0.02$ & 2.8 & +1.0 & $12.39\pm0.04$ & 2.3 & $6.26\pm0.72$ \\
 & +10.5 & $11.85\pm0.01$ & 4.9 & +10.5 & $11.85\pm0.02$ & 2.4 & $1.02\pm0.06$ \\
 & +21.1 & $11.23\pm0.04$ & 5.5 & \ldots & $<$10.0 & \ldots & $<$0.06 \\
\hline
\end{tabular}
\end{table*}

Details regarding the individual Na~{\sc i} and Ca~{\sc ii} components discerned through our profile fitting analysis are presented in Table~\ref{tab:components}. For each component, we list the LSR velocities, column densities, and $b$-values of Na~{\sc i} and Ca~{\sc ii}, along with the $N$(Na~{\sc i})/$N$(Ca~{\sc ii}) ratio. In cases where only Ca~{\sc ii} is detected, 3$\sigma$ upper limits on $N$(Na~{\sc i}) and $N$(Na~{\sc i})/$N$(Ca~{\sc ii}) are reported. Considering the resolution of our data, the velocities of the corresponding Na~{\sc i} and Ca~{\sc ii} components show good agreement. The median absolute difference in velocity between Na~{\sc i} and Ca~{\sc ii} is 0.9 km~s$^{-1}$. The Ca~{\sc ii} components are broader, with an average $b$-value of 4.5 km~s$^{-1}$ (identical to the velocity resolution). The average $b$-value of the Na~{\sc i} components is 2.6 km~s$^{-1}$. The Na~{\sc i}/Ca~{\sc ii} column density ratios exhibit a wide range. For components with Na~{\sc i} detections, the $N$(Na~{\sc i})/$N$(Ca~{\sc ii}) ratios range from $\sim$0.3 to $\sim$50, with higher ratios generally associated with larger Na~{\sc i} column densities. For components with Ca~{\sc ii} detections only, the upper limits on $N$(Na~{\sc i})/$N$(Ca~{\sc ii}) are $\sim$0.2 or less. Of particular note are the three high-velocity Ca~{\sc ii} components toward HD~198301, which exhibit very low upper limits on $N$(Na~{\sc i})/$N$(Ca~{\sc ii}).

\begin{table*}
\centering
\caption{Column densities (in cm$^{-2}$) in individual velocity components for the atomic and molecular species observed toward BD+31~4218. The velocities listed are the mean LSR velocities (in km~s$^{-1}$) averaged over all of the species in which a given component is detected. The CN column density refers to the total column density of CN in the $N=0$ and $N=1$ levels.}
\label{tab:atom_mol_comp}
\begin{tabular}{cccccccc}
\hline
$\langle v_{\rm LSR} \rangle$ & $\log N$(Ca~{\sc ii}) & $\log N$(Na~{\sc i}) & $\log N$(K~{\sc i}) & $\log N$(Ca~{\sc i}) & $\log N$(CH$^+$) & $\log N$(CH) & $\log N$(CN) \\
\hline
$-$18.7 & $10.81\pm0.04$ & \ldots & \ldots & \ldots & \ldots & \ldots & \ldots \\
$-$2.2 & $11.78\pm0.01$ & $12.98\pm0.06$ & $11.08\pm0.02$ & \ldots & $12.07\pm0.16$ & $12.17\pm0.11$ & \ldots \\
+4.1 & $12.48\pm0.03$ & $13.26\pm0.07$ & $11.35\pm0.02$ & $10.32\pm0.05$ & $12.69\pm0.05$ & $12.54\pm0.05$ & \ldots \\
+13.2 & $12.20\pm0.02$ & $13.88\pm0.07$ & $11.97\pm0.04$ & $10.05\pm0.06$ & \ldots & $13.33\pm0.01$ & $12.83\pm0.02$ \\
+22.1& $11.44\pm0.02$ & \ldots & \ldots & \ldots & \ldots & \ldots & \ldots \\
\hline
\end{tabular}
\end{table*}

Further details regarding the atomic and molecular component structure along the line of sight to BD+31~4218 are provided in Table~\ref{tab:atom_mol_comp}. The strongest Na~{\sc i} and K~{\sc i} component has a velocity of approximately +13 km~s$^{-1}$. This is also the strongest component in CH and the only component detected in CN absorption. The $N$(Na~{\sc i})/$N$(Ca~{\sc ii}) ratio for this component ($\sim$50) is the largest in our sample. The $N$(K~{\sc i})/$N$(Ca~{\sc i}) ratio for the +13 km~s$^{-1}$ component is also rather high ($\sim$80). Taken together, the high values for these two ratios indicate substantial depletion of Ca onto interstellar dust grains. The detection of strong CN and CH absorption, and the lack of CH$^+$ absorption at the same velocity, is a further indication of the presence of dense gas \citep[e.g.,][]{p05,s08}.

Three CN transitions are detected toward BD+31~4218: the $R$(0), $R$(1), and $P$(1) lines within the $B$$-$$X$ (0,~0) band near 3874~\AA{} (see Table~\ref{tab:totals}). To analyze these features, we used a modified version of the profile fitting routine, which performs a simultaneous fit to the three CN transitions, keeping the velocities and $b$-values consistent among the three lines. The column density in the $N=0$ rotational level is determined from the $R$(0) line, while the $N=1$ column density is derived from the $R$(1) and $P$(1) lines simultaneously. This procedure yields a $b$-value of 2.1 km~s$^{-1}$, similar to the $b$-values we find for Na~{\sc i} (2.4 km~s$^{-1}$) and K~{\sc i} (2.3 km~s$^{-1}$) for the same velocity component. The individual rotational column densities are $\log N(N=0)=12.67\pm0.02$ and $\log N(N=1)=12.33\pm0.03$. Thus, the rotational excitation temperature is $T_{01}({\rm CN})=2.90\pm0.15$~K. This result is consistent with excitation by the cosmic microwave background at a temperature of 2.725~K, but may also be indicative of a mild excess due to additional excitation by electron impact \citep[e.g.,][]{r11}.

\subsection{Comparison with the survey by Welsh et al.~(2002)}
The only previous high-resolution study of interstellar lines toward stars in the Cygnus Loop region is that of \citet{w02}. These authors studied Na~{\sc i} and Ca~{\sc ii} lines toward nine stars with projected on-sky positions within or near the Cygnus Loop SNR. At the time of their observations, the distances to the targets in \citet{w02} were not very well determined \citep[and the targets were selected assuming the distance to the Cygnus Loop was 440 pc;][]{b99}. Now, with the advent of accurate distances provided by the Gaia satellite, we can reevaluate the results of \citet{w02} and compare their results to those of our investigation. Most of the \citet{w02} targets have Gaia EDR3 distances that are less than $\sim$630 pc. These include HD~198597 ($213^{+2}_{-1}$ pc), HD~198056 ($310^{+3}_{-3}$ pc), HD~198197 ($392^{+7}_{-6}$ pc), HD~198946 ($445^{+13}_{-20}$ pc), HD~199102 ($536^{+8}_{-8}$ pc), HD~199042 ($594^{+10}_{-10}$ pc), and HD~335212 ($626^{+6}_{-7}$ pc). (Distances from \citet{bj21} are given in parentheses.) Only two of the \citet{w02} stars have distances greater than 700 pc. These are HD~198301 ($872^{+21}_{-20}$ pc), which is also a target in our investigation, and HD~197702 ($1072^{+31}_{-26}$ pc). All of our targets have Gaia EDR3 distances that are greater than $\sim$700 pc (Table~\ref{tab:targets}). Thus, by combining the \citet{w02} survey with our own, we can examine how the nature of the interstellar absorption changes as a function of distance toward the Cygnus Loop SNR.

The two closest stars in the \citet{w02} sample exhibit only a single velocity component in Na~{\sc i} and Ca~{\sc ii} at an LSR velocity near 0 km~s$^{-1}$, consistent with local foreground material. The \citet{w02} stars with distances greater than $\sim$350 pc generally exhibit multiple Na~{\sc i} and Ca~{\sc ii} components with average LSR velocities of +1, +9, +19, and +30 km~s$^{-1}$. These velocities are very similar to the velocities of the Na~{\sc i} and Ca~{\sc ii} components seen toward our targets (with the exception of the three high-velocity Ca~{\sc ii} components detected toward HD~198301). Since the range of absorption velocities is similar between the relatively nearby stars (at distances between $\sim$390 pc and $\sim$630 pc) and the stars that are further away (at distances greater than $\sim$700 pc), it is difficult to determine whether any of the low velocity components toward our stars might be associated with the Cygnus Loop SNR. At the longitude of the Cygnus Loop ($\sim$74\degr) and at the distance of the furthest star in our sample ($\sim$1100 pc), any gas participating in differential Galactic rotation would be expected to have an LSR velocity between 0 and +7 km~s$^{-1}$. Velocities well outside of this range might then be considered peculiar. For example, many of our targets exhibit gas components with velocities between +20 and +30 km~s$^{-1}$. However, while it might be tempting to ascribe this material to the effects of low velocity shocks associated with the Cygnus Loop, similar velocities are seen toward stars that are likely much closer than the SNR.

The only definitive conclusion we can reach in this regard is that the star HD~198301 must lie behind the Cygnus Loop SNR. Of the 14 stars in the combined sample of \citet{w02} and this paper, HD~198301 is the only star that exhibits truly high-velocity interstellar absorption. Indeed, our detection of Ca~{\sc ii} absorption approaching 100 km~s$^{-1}$ toward HD~198301 is the first conclusive detection of high-velocity, low-ionization gas associated with the Cygnus Loop. Curiously, HD~198301 was also observed by \citet{w02}. However, they make no mention of any high-velocity gas. (Their plot of the Ca~{\sc ii} spectrum in this direction extends to only $\pm$40 km~s$^{-1}$.) Thus, either the high-velocity features were not present in the spectrum obtained by \citet{w02} or those authors mistook these features for narrow stellar lines or for noise in the spectrum (B.~Welsh, 2023, private communication). Dramatic temporal variations in Na~{\sc i} and/or Ca~{\sc ii} lines have been seen toward numerous stars in the Vela SNR \citep[e.g.,][]{cs00,ra16,ra17,ra20} and toward one star in the Monoceros Loop \citep{dm16}. If the high-velocity features toward HD~198301 were not present in the spectrum that \citet{w02} obtained in 2001, then this would represent the first detection of temporal changes in interstellar absorption lines associated with the Cygnus Loop.

Finally, we note the similarity between the sight lines to the \citet{w02} target HD~197702 and our target BD+31~4218. Both stars are positioned beyond the northwestern boundary of the Cygnus Loop at distances in excess of 1000 pc. Like BD+31~4218, the interstellar Na~{\sc i} column density toward HD~197702 is more than an order of magnitude larger than toward any of the stars at distances less than 1000 pc. \citet{w02} reported a total Na~{\sc i} column density of $\log N({\rm Na~\textsc{i}})=14.36\pm0.22$ toward HD~197702, while we find $\log N({\rm Na~\textsc{i}})=14.02\pm0.07$ toward BD+31~4218. Both stars are evidently probing different portions of the western molecular cloud that \citet{f18b} suggested is physically interacting with the Cygnus Loop. Since the Na~{\sc i} column density toward BD+31~4224, a star positioned just 11\farcm6 away from BD+31~4218, is $\log N({\rm Na~\textsc{i}})=12.55\pm0.04$, the large jump in column density signifying the onset of this molecular cloud must occur at a distance between $\sim$730 pc and $\sim$1100 pc.

Tighter constraints on the distance to the western molecular cloud can be obtained by considering the 3D dust reddening map provided by \citet{g19}. \citet{f18a,f18b} have already performed a fairly extensive analysis of the variations in reddening with distance that are seen for clouds in the direction of the Cygnus Loop. Their analysis is based on the work of \citet{g15}, whereas we have used the updated dust map provided by \citet{g19}. The \citet{g19} data\footnote{Available from: \url{http://argonaut.skymaps.info/}} indicate that there is a large jump in $E$($B$$-$$V$) in the direction of BD+31~4218, from $\sim$0.06 to $\sim$0.3, for distances between 750 pc and 790 pc. A similar jump in reddening is indicated toward HD~197702. In that direction, $E$($B$$-$$V$) increases abruptly from $\sim$0.06 to $\sim$0.4 for distances between 790 pc and 840 pc. These limits are consistent with our less stringent constraints based on Na~{\sc i} absorption. In the next section, we discuss the implications of our results for distance estimates to the Cygnus Loop SNR.

\section{Discussion}
\citet{f21} recently reported a distance to the Cygnus Loop SNR of $725\pm15$ pc. This result, and its very small uncertainty, was based on the Gaia EDR3 distances to several stars that \citet{f18a,f18b} suggested are either interacting directly with the SNR or are positioned behind the expanding shock front. The star that \citet{f18a} claimed is directly interacting with the Cygnus Loop, due to the appearance of a bow-shock nebula surrounding the star and the chaotic nature of the SNR shocks in the vicinity of the star, is BD+31~4224, which has a Gaia EDR3 distance of $726^{+13}_{-11}$ pc. While there is qualitative support for the idea that the stellar wind from BD+31~4224 is interacting with the Cygnus Loop's expanding shocks \citep[see the discussion in][]{f18a}, the evidence is not conclusive. Another star considered by \citet{f21} in deriving their distance estimate is KPD~2055+3111, a subdwarf OB star that \citet{b09} reported shows high-velocity O~{\sc vi} absorption (at an LSR velocity of $-$75 km~s$^{-1}$). KPD~2055+3111 is positioned among the bright optical filamentary structures associated with the Eastern Veil Nebula. \citet{b09} based their analysis of this star on observations taken with the Far Ultraviolet Spectroscopic Explorer (FUSE). The FUSE spectrum of KPD~2055+3111 is rather complicated \citep[see Figure~7 in][]{b09}. However, if the detected O~{\sc vi} absorption is indeed associated with the SNR, then this implies that the distance to the Cygnus Loop is less than that to the star, which has a Gaia EDR3 distance of $819^{+21}_{-18}$ pc. (More precisely, since the high-velocity O~{\sc vi} absorption is blueshifted, this proves only that the background star is behind the approaching side of the SNR shock front.)

The three other stars that \citet{f21} considered in deriving their precision distance estimate to the Cygnus Loop are those that \citet{f18b} claimed show high-velocity Na~{\sc i} and Ca~{\sc ii} absorption. The most prominent among these (because it appears to show the clearest example of redshifted and blueshifted Na~{\sc i} and Ca~{\sc ii} absorption) is HD~335334 (referred to in \citet{f18b,f21} as Star X). However, as we have shown conclusively in this paper, HD~335334 is a double-line spectroscopic binary and shows no evidence for high-velocity interstellar absorption. As such, the distance to this star cannot be used to constrain the distance to the Cygnus Loop. The other two stars are TYC~2688-365-1 (Star Y), and TYC~2692-3378-1 (Star Z). We did not observe these stars with the 2.7 m telescope (because they are somewhat too faint for high-resolution, high S/N ratio spectroscopy). However, from the spectra of these stars presented in \citet{f18b}, the ``high-velocity'' Na~{\sc i} features appear instead to be broad, stellar Na~{\sc i} absorption lines onto which the narrow low-velocity interstellar Na~{\sc i} components are superimposed. \citep[The Ca~{\sc ii} feature in the spectrum of Star Z is clearly a stellar absorption line; see Figure~3 in][]{f18b}.

One implication of these results involves the detailed morphological orientation of the Cygnus Loop along the line of sight. \citet{f18b} struggled to obtain a distance estimate that simultaneously met the criteria that BD+31~4224 be inside the remnant and the seemingly closer stars X and Y be behind the remnant. \citet{f18b} proposed a solution in which the Cygnus Loop's main northern shell is aspherical and tilted so that these different criteria could be accomodated \citep[see Figures~4 and 5 in][]{f18b}. Since neither Star X nor Star Y show evidence of high-velocity interstellar absorption, there is no longer any need for this complicated ``solution.'' \citet{f21} pointed to an additional tension that arises with their restrictive distance estimate of only 725 pc. Proper motion measurements of several of the Cygnus Loop's northern nonradiative Balmer-dominated filaments by \citet{s09}, combined with a distance of 725 pc, yield shock velocities that are too low compared to the shock velocities determined from line-width measurements of the broad H$\alpha$ emission components at the same filamentary positions \citep{m14}. From their H$\alpha$ emission measurements, \citet{m14} suggested that a more likely distance to the Cygnus Loop SNR is $\sim$890 pc. (A reanalysis by \citet{r15}, which considered the effects of thermal equilibration in a collisionless shock, reduced this distance estimate to $\sim$800 pc.)

Our discovery of high-velocity Ca~{\sc ii} absorption (at LSR velocities of +62, +82, and +96 km~s$^{-1}$) toward HD~198301 implies that the distance to the Cygnus Loop SNR must be less than the distance to this star, which has a Gaia EDR3 distance of $872^{+21}_{-20}$ pc. Note that the fact that the high-velocity gas is redshifted means that the star must be behind the receding edge of the SNR shock front. The star HD~198301 is positioned in the midst of a bright triangular-shaped filamentary region known as Pickering's Triangle\footnote{``Pickering's Triangle'' was discovered photographically in 1904 by Williamina Fleming, an astronomer working at the Harvard College Observatory under the direction of Edward Charles Pickering \citep[see][]{pf06}.} (see Figure~\ref{fig:cygnus_dss2}). The bright optical emission from this region (along with the similarly bright filamentary emission from the Eastern and Western Veil Nebulae) has been interpreted as arising from the interaction of the SNR blast wave with density inhomogeneities in the surrounding interstellar clouds \citep{l98,f18b}. Since the high-velocity shocked gas toward HD~198301 is redshifted, this strongly implies that the density inhomogeneity giving rise to the optical emission associated with Pickering's Triangle lies on the rear side of the expanding SNR shock front. (A similar argument implies that the shocked gas associated with the Eastern Veil Nebula lies on the near side of the SNR, since the high-velocity O~{\sc vi} absorption toward KPD~2055+3111 is blueshifted.)

Another implication of our results is that the magnitude of the velocities of shocks driven into interstellar clouds by the Cygnus Loop's blast wave is less than that implied by the analysis of \citet{f18b}. Those authors reported Na~{\sc i} components with velocities ranging from $-$160 to +240 km~s$^{-1}$ (toward Star Z). However, as discussed above, these supposed high-velocity ``components'' are more likely just the extreme portions of the wings of broad stellar Na~{\sc i} absorption lines. The maximum Ca~{\sc ii} absorption velocity toward HD~198301 (+96 km~s$^{-1}$) implies a cloud shock velocity of $\sim$100 km~s$^{-1}$, much less than the 240 km~s$^{-1}$ reported by \citet{f18b}. The velocity of the cloud shock toward HD~198301 could be somewhat higher than 100 km~s$^{-1}$ if there is a significant tangential component to the motion. Nevertheless, our determination of the (radial component of the) cloud shock velocity toward HD~198301 is consistent with the shock velocities derived from proper motion and emission-line studies of the bright radiative filaments associated with the Cygnus Loop, which indicate shock velocities in the range 100 to 150 km~s$^{-1}$ \citep[e.g.,][]{ray20}.

High-velocity interstellar absorption features are relatively rare, even among sight lines passing through the optical boundaries of SNRs. From a large survey of Na~{\sc i} and Ca ~{\sc ii} lines toward stars in the Vela SNR, \citet{cs00} found that only $\sim$25\% of the stars showed high-velocity Ca~{\sc ii} components, and only $\sim$12\% showed high-velocity Na~{\sc i}. Of the 14 stars in the Cygnus Loop region included in \citet{w02} and in this paper, only one exhibits high-velocity Ca~{\sc ii} absorption. Partly, this is due to a distance effect since many of the \citet{w02} targets are likely positioned in front of the SNR. However, our targets were deliberately chosen because they are more likely to be background stars, yet only one of our six targets shows high-velocity interstellar absorption. In this context, it is important to remember that low-ionization species, such as Na~{\sc i} and Ca~{\sc ii}, do not probe the SNR shock itself. Rather, they probe pre-existing interstellar gas that has been shocked and accelerated by the SNR blast wave. Thus, the detection of high-velocity interstellar Na~{\sc i} or Ca~{\sc ii} absorption requires the chance alignment of a shocked interstellar cloud and a bright background star. Such chance alignments are evidently somewhat rare given the highly inhomogeneous nature of the interstellar medium.

\section{Summary and Conclusions}
Six stars were observed at moderately high spectral resolution with the Tull spectrograph and 2.7 m telescope at McDonald Observatory. Low velocity interstellar Na~{\sc i} and Ca~{\sc ii} absorption lines are detected in each direction. High-velocity Ca~{\sc ii} absorption (at LSR velocities of +62, +82, and +96 km~s$^{-1}$) is detected toward only one star: HD~198301, which lies behind the bright region of filamentary emission known as Pickering's Triangle. This is the first conclusive detection of high-velocity gas in a low ionization species such as Ca~{\sc ii} associated with the Cygnus Loop SNR. This detection means that the receding edge of the Cygnus Loop's shock front must be in front of HD~198301 (which is at a distance of $\sim$870 pc). A previous detection of high-velocity O~{\sc vi} absorption toward KPD~2055+3111 (at $-$75 km~s$^{-1}$) by \citet{b09} indicates that the approaching side of the shock front must be at a distance of less than $\sim$820 pc. While \citet{f21} constrained the distance to the Cygnus Loop to be $725\pm15$ pc, this result is largely dependent on whether or not the star BD+31~4224 is physically interacting with the Cygnus Loop. However, the evidence suggesting an interaction between this star's stellar wind and the Cygnus Loop's expanding shock wave is merely suggestive and not conclusive.

The star HD~335334, which was previously thought to exhibit high-velocity Na~{\sc i} and Ca~{\sc ii} absorption \citep{f18b}, is actually a double-line spectroscopic binary star. We find no evidence for high-velocity interstellar absorption in this direction, meaning that the distance to HD~335334 cannot be used to constrain the distance to the Cygnus Loop. Two other stars observed by \citet{f18b} probably also do not exhibit high-velocity interstellar absorption. The ``high-velocity'' Na~{\sc i} components in these directions are instead portions of the broad stellar Na~{\sc i} absorption lines. The end result of our analysis is that the distance to the Cygnus Loop SNR is not as precisely known as \citet{f21} have claimed.

Strong interstellar absorption from various atomic and molecular species is detected toward the most distant star in our sample: BD+31~4218, which is positioned to the northwest beyond the bright optical boundary of the Cygnus Loop SNR. This star probes part of an adjacent molecular cloud to the west of the Cygnus Loop. The rear portion of the expanding SNR shock wave appears to be directly interacting with this molecular material, giving rise to the Western Veil Nebula and probably also Pickering's Triangle. The rise in column density associated with this molecular material is constrained to lie between $\sim$730 and $\sim$1100 pc. A corresponding jump in $E$($B$$-$$V$) in the direction of BD+31~4218 is indicated for distances between 750 and 790 pc \citep{g19}. If the Cygnus Loop SNR is indeed interacting with the background molecular cloud to its west, then the distance to the SNR would likely need to fall within this range. We note that the original \citet{m58} value of 770 pc would fit within this constraint.

The physical conditions in the high-velocity shocked material toward HD~198301 could be studied in much greater detail using high-resolution HST/STIS observations in the UV. The UV portion of the spectrum provides access to numerous diagnostic lines that can be used to examine the densities, temperatures, pressures, depletions, and ionization states of shocked interstellar gas \citep[e.g.,][]{rjf20}. However, because HD~198301 is a narrow-lined star, its UV spectrum will be challenging to interpret. Nevertheless, this star provides us with the best (and, indeed, only) opportunity to examine the detailed physical conditions in high-velocity shocked gas associated with the Cygnus Loop SNR using the technique of high-resolution UV absorption-line spectroscopy.

\section*{Acknowledgements}
We thank Coyne Gibson of McDonald Observatory for his help in manually aligning the Tull spectrograph in its TS23 configuration. This research has made use of the SIMBAD database, operated at CDS, Strasbourg, France.

\section*{Data Availability}
The McDonald Observatory data on which our analysis is based may be available upon request to the corresponding lead author.
 







\appendix

\section{Stellar Spectral Classification}
Before this investigation, most of our targets had very little reliable information available regarding their spectral types and luminosity classes. This was especially problematic for HD~198301 and HD~335334. Both stars exhibit narrow stellar absorption lines, making the task of distinguishing between stellar absorption and interstellar absorption challenging. For this reason, and because our spectra cover nearly the entire optical range, we sought to obtain more accurate information regarding the spectral classification of our stars.

\begin{figure*}
\includegraphics[width=\textwidth]{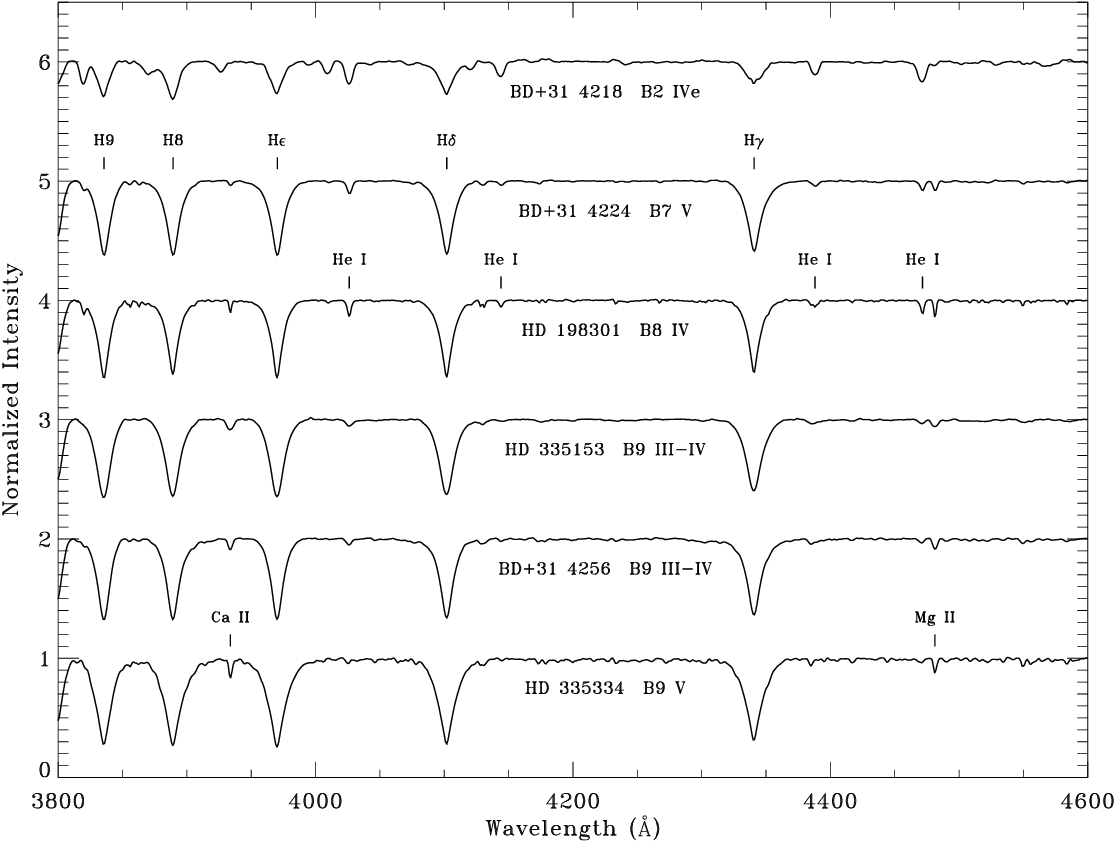}
\caption{Heavily smoothed spectra of the program stars used for stellar classification. The observed spectra have been normalized to the continuum, smoothed to a resolution of 1.8~\AA{}, and rebinned to a spacing of 1~\AA{}. The spectra are offset from one another for clarity. Prominent stellar absorption features used in the classification process are labelled.}
\label{fig:cygnus_atlas}
\end{figure*}

The first step toward classifying our targets was to create classification grade spectra from our high-resolution echelle data. By ``classification grade'' we mean low-resolution, continuum-normalized spectra in the blue-violet region ($\sim$3800--4600~\AA{}). The individual echelle orders from our high-resolution data were carefully normalized and merged into a continuous spectrum. Then, using computer programs associated with the spectral classification routine MKCLASS\footnote{Available from: \url{http://www.appstate.edu/~grayro/mkclass/}} \citep{gc14}, the normalized spectra were smoothed to a resolution of 1.8~\AA{} and rebinned to a spacing of 1 \AA{}. The resulting classification grade spectra are shown in Figure~\ref{fig:cygnus_atlas}.

Initial spectral types were determined by comparing the heavily smoothed spectra of our targets with the library of standard spectra provided with the MKCLASS program. More quantitative results were then obtained by measuring the equivalent widths of prominent absorption lines in our program stars and comparing our measurements to similar measurements made for spectral-type standards \citep{m13}. In particular, we measured equivalent widths for Ca~{\sc ii}~$\lambda3933$, He~{\sc i}~$\lambda4009$, He~{\sc i}~$\lambda4026$, H$\delta$~$\lambda4101$, He~{\sc i}~$\lambda4143$, H$\gamma$~$\lambda4340$, He~{\sc i}~$\lambda4471$, Mg~{\sc ii}~$\lambda4481$, H$\beta$~$\lambda4862$, and He~{\sc i}~$\lambda4921$. Since the He~{\sc i} lines weaken, while lines such as Ca~{\sc ii}~$\lambda3933$ and Mg~{\sc ii}~$\lambda4481$ strengthen, with decreasing temperature for late B into early A stars \citep{gc09}, the He~{\sc i}~$\lambda4471$/Mg~{\sc ii}~$\lambda4481$ equivalent width ratio was especially useful for deriving temperature types for our stars. Luminosity classes were then obtained by comparing the detailed shapes of the H Balmer lines to those of the standard stars from the MKCLASS library.

The final derived spectral types and luminosity classes for our targets are provided in Table~\ref{tab:targets}. One of our targets, BD+31~4218, is a Be star, with prominent double-peaked emission in many of the H Balmer lines, including H$\alpha$, H$\beta$, H$\gamma$, and H$\delta$. It was more difficult to derive a luminosity class for this star since the shapes (and absorption strengths) of the Balmer lines are modified by emission. However, we found that the spectrum of BD+31~4218 closely resembled that of the B2 IVpne star HD~88661 \citep[see Figure~4.17 in][]{gc09}, although our target exhibits somewhat less emission in the Balmer lines than does HD~88661. Our equivalent width analysis had already yielded a spectral type of B2 for BD+31~4218. We therefore adopted the luminosity class from HD~88661 for BD+31~4218.

\begin{table}
\centering
\caption{Best-fitting model parameters for the program stars (excluding the Be star BD+31~4218) determined by comparing the observed spectra to the high-resolution synthetic spectra of \citet{m05}. We list model parameters for both components of the spectroscopic binary star HD~335334.}
\label{tab:models}
\begin{tabular}{lccc}
\hline
Star & $T_{\rm eff}$ & $\log g$ & $v \sin i$ \\
 & (K) & & (km~s$^{-1}$) \\
\hline
HD~335153 & 11,000 & 3.5 & 250 \\
BD+31~4224 & 13,000 & 4.0 & 200 \\
HD~198301 & 12,000 & 3.5 & 20 \\
BD+31~4256 & 11,000 & 3.5 & 200 \\
HD~335334A & 10,500 & 3.5 & 20 \\
HD~335334B & 10,500 & 4.0 & 20 \\
\hline
\end{tabular}
\end{table}

We checked the accuracy of our derived spectral types by comparing the unsmoothed (high-resolution) spectra of our targets to the library of high-resolution stellar model spectra provided by \citet{m05}. The \citet{m05} library provides a maximum resolving power of $R=20,000$, which is not quite as high as the resolving power achieved with our spectra ($R\approx66,000$). Nevertheless, comparison with the \citet{m05} models proved useful, especially for the two stars in our sample with narrow absorption lines (see Section 3.1). The grid of \citet{m05} models for late B stars has steps in $T_{\rm eff}$ of 500 K or 1,000 K and steps in $\log g$ of 0.5 dex. We considered solar metallicity models only. For a given star, we compared the observed high-resolution spectrum to a grid of models with $T_{\rm eff}$ and $\log g$ close to that of the derived spectral type and luminosity class. We also tested several different values of the projected rotational velocity. The derived best-fitting values of $T_{\rm eff}$, $\log g$, and $v \sin i$ are provided in Table~\ref{tab:models}. We list model parameters for both components of the spectroscopic binary star HD~335334. These are the parameters that were used to construct the composite spectra discussed in Section~3.1.2. The star BD+31~4218 is not included in Table~\ref{tab:models} because it would not be appropriate to compare the observed spectrum of this Be star with the model spectra provided by \citet{m05}. In general, the best-fitting model parameters listed in Table~\ref{tab:models} help to confirm the spectral types and luminosity classes previously derived for our program stars.


\bsp	
\label{lastpage}
\end{document}